\documentclass[5p]{elsarticle}
\journal{Technical Report}
\usepackage{amsmath}
\newtheorem{theorem}{\textbf{Theorem}}
\newtheorem{lemma}{\textbf{Lemma}}
\newtheorem{example}{\textbf{Example}}
\newtheorem{corollary}{\textbf{Corollary}}
\newtheorem{remark}{\textbf{Remark}}
\newtheorem{definition}{\textbf{Definition}}

\newtheorem{proposition}{\textbf{Proposition}}

\usepackage{multirow} %multirow for format of table
\usepackage{xcolor}
\usepackage{multirow}
\usepackage{setspace}
\usepackage{geometry}
\usepackage{url}
\usepackage{subfigure}
\usepackage{fancyhdr}
\usepackage{amsmath}
\usepackage{multirow}
\usepackage{amssymb }
\usepackage{color}
\usepackage{graphics} % for pdf,  bitmapped graphics files
\usepackage{graphicx}
\usepackage{algorithm} %format of the algorithm
\usepackage{algorithmic} %format of the algorithm
\usepackage{subeqnarray}
\usepackage{cases}

\newcommand\myeqa{\mathrel{\stackrel{\makebox[0pt]{\mbox{\normalfont\tiny (a)}}}{=}}}
\newcommand\myineqa{\mathrel{\stackrel{\makebox[0pt]{\mbox{\normalfont\tiny (a)}}}{\ne}}}
 \normalsize
\begin{document}

\begin{frontmatter}

\title{Partial Strong Structural Controllability } %of Linear Systems
%Controllability of Networked Relative Coupling Systems: a Structural Analysis

%\author{Yuan Zhang and Tong Zhou$^{\dag}$% <-this % stops a space
%\thanks{*This work was supported in part by the NNSFC under Grant 61573209 and 61733008. {\bf This work is just an extension of the conference paper \cite{Y_Zhang_2017}}.}% <-this % stops a space
%\thanks{$^{\dag}$Yuan Zhang and Tong Zhou are with the Department of Automation and TNList, Tsinghua University, Beijing, 100084, P.~R.~China
%        {(email: {\tt\small zhangyuan14@mails.tsinghua.edu.cn, tzhou@mail.tsinghua.edu.cn}).}}%
%}

\author{Yuan Zhang, Yuanqing Xia}%, Jinhui Zhang
\address{Department of Automatic Control, Beijing Institute of Technology, Beijing, China\\~Email: $\emph{\{zhangyuan14,xia\_yuanqing\}@bit.edu.cn}$} %,zhangjinh
%\thanks{This work was supported in part by the China Postdoctoral Innovative Talent Support Program under Grant BX20200055,  the China Postdoctoral
%Science Foundation under Grant 2020M680016, and the National Natural Science Foundation of China under Grant 62003042. Corresponding author: Y. Xia}

%\ead{{zhangyuan14@bit.edu.cn,xia_yuanqing@bit.edu.cn@bit.edu.cn}}
\fntext[myfootnote]{This work was supported in part by the China Postdoctoral Innovative Talent Support Program under Grant BX20200055,  the China Postdoctoral
Science Foundation under Grant 2020M680016, and the National Natural Science Foundation of China under Grant 62003042. Corresponding author: Y. Xia.} %zhangyuan14@bit.edu.cn, xia_yuanqing@bit.edu.cn@bit.edu.cn}
{\small
\begin{abstract}This paper introduces a new controllability notion, termed partial strong structural controllability (PSSC), on a structured system whose entries of system matrices are either fixed zero or indeterminate, which naturally extends the conventional strong
structural controllability (SSC) and bridges the gap between structural controllability and SSC. Dividing the indeterminate
entries into two categories, generic entries and unspecified entries, a system is PSSC, if for almost all values of the generic
entries in the parameter space except for a set of measure zero, and any nonzero (complex) values of the unspecified
entries, the corresponding system is controllable. We highlight that this notion generalizes the
generic property embedded in the conventional structural controllability for single-input systems. We then give algebraic and (bipartite) graph-theoretic necessary and sufficient conditions for single-input systems to be PSSC. Conditions for multi-input systems are subsequently given for a special case. It is shown the established results can induce a new maximum matching based criterion
for SSC over the system bipartite graph representations.
\end{abstract}

}
\end{frontmatter}
%\keywords{Controllability robustness, actuator removals, security analysis, recursive tree, submodular function minimization, complexity}

{\small
%%%%%%%%%%%%%%%%%%%%%%%%%%%%%%%%%%%%%%%%%%%%%%%%%%%%%%%%%%%%%%%%%%%%%%%%%%%%%%%%
\section{Introduction} \label{intro-sec}
The past decades have witnessed an explosion of research interest into control and observation of complex networks \cite{Y.Y.2011Controllability,liu2013observability,yan2012controlling}. This is perhaps because many real-world systems could be naturally modeled as complex dynamic networks, such as social networks, biological networks, traffic networks, as well as the internet \cite{barabasi1999emergence}. Among the related problems, network controllability has been extensively explored both from a qualitative \cite{klamka1963controllability,aguilar2015graph,Y_Zhang_2016} and quantitative perspective \cite{muller1972analysis,yan2012controlling,P.Fa2014Controllability}. %The former mainly seeks to understand how the network structure and the nodal dynamics influence the system performance in terms of the classical Kalman controllability, e.g., \cite{R.Am2009Controllability,notarstefano2013controllability,aguilar2015graph,Y_Zhang_2016}. The latter usually adopts some controllability Gramian-related metrics to characterize the control energy required for a control task, e.g., \cite{yan2012controlling,P.Fa2014Controllability}. %, as well as the asymptotic behaviors of the required control energy versus the number of control nodes, Whiles

A well-accepted alternative of the classical Kalman controllability is the notion of {\emph{structural controllability}}. This notion was first introduced by Lin \cite{Lin_1974} by inspecting that controllability is a generic property, in the sense that depending on the structure of the state-space matrices, either the corresponding system realization is controllable for almost all values of the indeterminate entries, or there is no controllable system realization. Various criteria for structural controllability were found subsequently \cite{shields1976structural,mayeda1981structural}. Structural controllability is promising from a practical view, since it does not require the accurate values of system parameters, thus immune to numerical errors. Moreover, criteria for structural controllability usually are directly linked to some mild interconnection conditions of the graph associated with the system structure \cite{generic}, making it attractive in analyzing large-scale network systems \cite{Y.Y.2011Controllability,Composability,zhang2019structural}. %zero/nonzero pattern

A more stringent notion than structural controllability is the strong structural controllability (SSC), which requires the system to be controllable for {\emph{all}} (nonzero) values of the indeterminate entries \cite{strong_controllability}. As argued in \cite{strong_controllability}, although the uncontrollable system realizations are atypical if a system is structurally controllable, there do exist scenarios where the existence of an uncontrollable system realization is prohibited, especially in some critical infrastructures where high-level robustness of the system controllability is required. The first criterion for SSC of single-input systems was given in \cite{strong_controllability}, followed by some graph-theoretic characterizations of SSC of multi-input systems in \cite{bowden2012strong,jarczyk2011strong}. \cite{chapman2013strong} provided a constrained-matching based criterion, while \cite{monshizadeh2014zero} related SSC to the zero-forcing set and graph-coloring. Further, SSC of undirected networks was studied in \cite{mousavi2018structural}. For a comprehensive comparison of these criteria, see \cite{jia2020unifying}.

\begin{figure}
  \centering
  \includegraphics[width=1.68in]{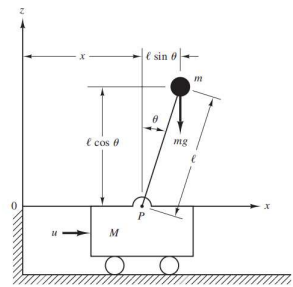}\\
  \caption{Inverted-pendulum system. Figure borrowed from \cite{Modern_Control_Ogata}.}\label{inverted-pendulum}
\end{figure}

It is well-accepted that being SSC requires a system to have a much more restrictive structure than being structurally controllable. In fact, it is shown in \cite{godsil2012controllable,o2016conjecture} that, for structural controllability, the ratio of controllable graphs (networks) to the total number of graphs with $n$ nodes tends to one, as $n\to \infty$. By contrast, \cite{menara2017number} demonstrates that for SSC, the same ratio approaches zero, provided all nodes have self-loops and the ratio of control nodes to $n$ converges to zero. This phenomenon indicates that there is an essential gap between structural controllability and SSC. On the other hand, from a generic view, SSC requires that the system realization is controllable subject to {\emph{all}} possible interrelations among the (nonzero) indeterminate parameters.
In contrast, structural controllability only requires the system realization to be controllable assuming independence among all the indeterminate parameters. Note in practice, it is possible that only a partial subset of indeterminate parameters may be interrelated while the rest are independent of each other. This means, even a system itself is not SSC, the robustness requirement for system controllability might be met.  The following example illustrates this.

\begin{example}[Motivating example] \label{examp-motivate} Consider the following \\~ inverted-pendulum system shown in Fig. \ref{inverted-pendulum}, which also appears in \citep[Example 3--6]{Modern_Control_Ogata}. Let $\theta$ and $x$ be respectively the angle of the pendulum rod and the location of the cart. Define state variables $x_1,x_2,x_3$, and $x_4$ by $x_1=\theta$, $x_2=\dot \theta$, $x_3=x$, $x_4=\dot x$. According to \cite{Modern_Control_Ogata}, the linearized state-space equation of this system reads
$${\small\left[
  \begin{array}{c}
    \dot x_1 \\
    \dot x_2 \\
    \dot x_3 \\
    \dot x_4 \\
  \end{array}
\right]=\underbrace{\left[
          \begin{array}{cccc}
            0 & 1 & 0 & 0\\
            \frac{(M+m)g}{Ml} & 0 & 0 & 0 \\
            0 & 0 & 0 & 1 \\
            -\frac{mg}{M} & 0 & 0 & 0 \\
          \end{array}
        \right]}_{A=[a_{ij}]}\left[
                 \begin{array}{c}
                   x_1 \\
                   x_2 \\
                   x_3 \\
                   x_4 \\
                 \end{array}
               \right]+\underbrace{\left[
                         \begin{array}{c}
                           0 \\
                           -\frac{1}{Ml} \\
                           0 \\
                           \frac{1}{M} \\
                         \end{array}
                       \right]}_{B=[b_{ij}]}u}$$
where $m$ and $M$ are the masses of the pendulum bob and the cart, $l$ is the length of the massless rod, $u$ is the force imposed on the cart, and $g$ is the acceleration of gravity. From \cite{strong_controllability}, the zero/nonzero structure of $(A,B)$ is not SSC. However, direct calculation shows that the determinant of the controllability matrix of $(A,B)$ is $a_{12}^3a_{34}b_{21}^2(a_{21}b_{41} - a_{41}b_{21})^2$. This means, if we regard $a_{21}, a_{41}, b_{41}$, and $b_{21}$ as independent parameters, leading to the usual satisfaction of $a_{21}b_{41} - a_{41}b_{21}\ne 0$, then whatever nonzero values $a_{12}$ and $a_{34}$ may take, the corresponding system will be controllable. In fact, back to this example, $a_{21}b_{41} - a_{41}b_{21}= {\footnotesize {\frac{Mg}{M^2l}}}\ne 0$ is always fulfilled. \hfill $\square$
\end{example}

Motivated by the above observations, this paper proposes a new controllability notion, named partial strong structural controllability (PSSC), trying to bridge the gap between structural controllability and SSC. In this notion, the indeterminate entries of a (structured) system are divided into two categories, generic entries that are assumed to take independent values, and unspecified entries that can take arbitrary nonzero (complex) values. A system is PSSC, if for {\emph{almost all}} values of the generic entries, the corresponding system realization is controllable for {\emph{all}} nonzero values of the unspecified entries. The main contributions are detailed as follows:

{
1) We propose a novel controllability notion of PSSC for linear structured systems. This notion naturally extends the conventional SSC, and for single-input systems, inherits and generalizes the generic property characterized by the conventional structural controllability.

2) We give an algebraic and a graph-theoretic necessary and sufficient conditions for single-input systems to be PSSC,  the latter of which can be verified in polynomial time.

3) We extend 2) to the multi-input systems subject to the single unspecified entry constraint.

4) Finally, it is shown the established results can provide a new graph-theoretic criterion for SSC of single-input systems involving maximum matchings over the system bipartite graph representations.} %maximum  bipartite matchings.}  %

%4) We also extend PSSC to the case where the unspecified entries can take either nonzero values or arbitrary values. We reveal a separation principle to
%deal with such a mix of entries for single-input systems. %(including zero and nonzero)

%5) Finally, we demonstrate that the established results can provide a new graph-theoretic criterion for SSC in terms of maximum matchings of the system bipartite graph representations, and relate it to one existing criterion in \cite{chapman2013strong} for SSC of real systems. %by showing its equivalence to one existing criterion in \cite{chapman2013strong} for SSC of real systems.

% To the best of our knowledge, this is the first work to build a bridge between structural controllability and SSC, which, notably,

{Notably, our work provides a unifying viewpoint towards two seemingly different controllability notions: structural controllability and SSC.} It is worth noting that a related notion is the perturbation-tolerant structural controllability proposed in \cite{full-version-tac}. Compared to that notion, PSSC addressed in this paper is an essentially different concept with many different properties. Particularly, PSSC contains the conventional SSC as a special case, while PTSC does not. Furthermore, from a technical view, due to the nonzero constraint of the unspecified entries, the techniques in \cite{full-version-tac} are not sufficient to characterize PSSC, and it is nontrivial to extend them to obtain the presented criteria here. %it is nontrivial to extend the technique line in \cite{full-version-tac} to obtain the presented conditions for PSSC. %

The rest of this paper is organized as follows. Section \ref{problem-formulation} presents the problem formulation and some preliminaries. Section \ref{single-case} justifies the generic property involved in PSSC and provides necessary and sufficient conditions for single-input systems to be PSSC. Section \ref{multi-case} extends these results to the multi-input case subject to the single unspecified entry constraint.  Section \ref{SSC-corollary-section} discusses the implications of the established results for the existing SSC theory. All proofs of the technical results are given in the appendix.

%In Section \ref{extension-sec}, PSSC is extended to including the unspecified entries that can take arbitrary values. All proofs of the technical results are given in the appendix.

% provides useful insights to the PSSC addressed in this paper, but is not sufficient to deal with the nonzero weight constraint. Particularly, the criteria for PTSC do not contain those for SSC as special cases, while it is the case for PSSC. We have carefully proved the equivalence between our criteria and the existing ones for SSC. Nevertheless, some intermediate results in \cite{full-version-tac} are adopted in this paper as auxiliary results, especially those on zeros of generic matrix pencils.  Our criteria also provide a new sight on SSC, namely, using maximum (weighted) matching. To the best of our knowledge, the existing criteria for SSC are based on the zero-forcing set, color change rule or other tools \cite{jia2020unifying,mousavi2018structural}, but not including maximum matching.\footnote{The so-called constrained matching condition in \cite{chapman2013strong} is used to check and find SSC input sets, and the input has a specific structure. }

{\bf Notations:} ${\mathbb N}, {\mathbb R}, {\mathbb C}$ denote the sets of natural, real, and complex numbers, respectively. Let $\bar {\mathbb R}={\mathbb R}\backslash \{0\}$, $\bar {\mathbb C}={\mathbb C}\backslash \{0\}$, and $\bar {\mathbb C}^{n}$ (resp. $\bar {\mathbb R}^{n}$) be the set of $n$-dimensional complex (real) vectors with each of its entries being {\emph{nonzero}}. For $n\in {\mathbb N}$, $J_n$ stands for $\{1,...,n\}$. For an $n\times m$ matrix $M$, $M[I_1,I_2]$ denotes the submatrix of $M$ whose rows are indexed by $I_1\subseteq J_n$ and columns by $I_2\subseteq J_m$.

%Define the set $\bar {\mathbb C}^{n_\times}$ as
%$$\bar {\mathbb C}^{n_\times}=\{\bar p\in {\mathbb C}^{n_\times}: \bar p_i\ne 0, i=1,...,n_\times\}. $$

\section{Problem formulation and preliminaries} \label{problem-formulation}
\subsection{Structured matrix and generic matrix} \label{structured-matrix-definition}
We often use a structured matrix to denote the sparsity pattern of a numerical matrix \cite{Murota_Book}. A structured matrix is a matrix whose entries are chosen from the set $\{0,*,\times\}$. The $*$ entries and $\times$ entries are both called indeterminate entries. We use $\{0,*,\times\}^{p\times q}$ to denote set of all $p\times q$ structured matrices with entries from $\{0,*,\times\}$. Moreover, for $\bar M \in \{0,*,\times\}^{p\times q}$, ${\bf S}_{\bar M}$ is defined as
\[{\bf S}_{\bar M}\!=\!\{M\in {\mathbb C}^{p\times q}: M_{ij}\ne 0 \ {\text {if}}\ \bar M_{ij}=\times, M_{ij}=0 \ {\text {if}}\ \bar M_{ij}=0\}.\]
That is, a $*$ entry can take either zero or nonzero, while a $\times$ entry can take only nonzero value.  An $M\in {\bf S}_{\bar M}$ is called a realization of $\bar M$. We also define two structured matrices $\bar M^*\in \{0,*\}^{p\times q}$ and $\bar M^\times\in \{0,\times\}^{p\times q}$ associated with $\bar M \in \{0,*,\times\}^{p\times q}$ respectively as
{\small\begin{equation}\label{star-matrix-def} \bar M_{ij}^*=\left\{ \begin{aligned}
&= *, {\text {if}} \ \bar M_{ij}=\times \\
&= \bar M_{ij}, {\text{ else}}
\end{aligned}\right., \ \bar M_{ij}^\times =\left \{ \begin{aligned}
&= \times, {\text {if}}\ \bar M_{ij}= * \\
&= \bar M_{ij}, {\text{ else}}.
\end{aligned}
\right.  \end{equation}}That is, $\bar M^*$ (resp. $\bar M^\times$) is obtained from $\bar M$ by replacing all of its indeterminate entries with $*$ (resp. $\times$) entries.

The operation `+' between two structured matrices with the same dimension is an entry-wise addition operation so that every indeterminate entry in the sum appears exactly once in one of the addends. We define a {\emph{generic realization}} of $\bar M$ as the realization whose indeterminate entries are assigned with independent parameters, and call such a realization a generic matrix without specifying the corresponding structured matrix. The {\emph{generic rank}} of a generic matrix (or structured matrix), denoted by ${\rm grank}(\cdot)$, is the maximum rank it achieves as the function of its indeterminate parameters. For a generic matrix $M$ and a constant matrix $N$ with the same dimension, $M+\lambda N$ defines a {\emph{generic matrix pencil}} \cite{Murota_Book}, which is a matrix-valued polynomial of the independent parameters in $M$ and the variable $\lambda$.

%We pay special attention to the typical realization of a structured matrix. To this end,

\subsection{Notion of PSSC}
Consider a linear time invariant system as
\begin{equation}
\label{plant}
\dot x(t)= A x(t)+B u(t),
\end{equation}
where $x(t)\in {\mathbb C}^n$ is the state variable, $u(t)\in {\mathbb C}^m$ is the input, $A\in {\mathbb C}^{n\times n}$ and $B\in {\mathbb C}^{n\times m}$ are respectively the state transition matrix and input matrix. For description simplicity, by the pair $(A,B)$ we refer to a system described by (\ref{plant}) that can be either single-input (i.e., $m=1$) or multi-input (i.e., $m>1$), while by $(A,b)$ we refer to a single-input system. {Note dynamic systems with complex-valued system matrices and state variables are not rare in the literature \cite{R.E1984Between,gu2000new,mengi2008estimation,wu2016adaptive,lin2014distributed}, either for practical utility or for theoretical studies. Following a similar manner as $(A,B)$ in the real field \citep[Chap. 9]{Modern_Control_Ogata}, it turns out the existence and uniqueness of solutions of system (\ref{plant}) are guaranteed even in the complex domain. }
% from which we know system (\ref{plant}) is well-posed even in the complex domain \cite{Modern_Control_Ogata}. } %It is evident that the system solutions of system (\ref{plant}) is guaranteed.

Denote the controllability matrix of system (\ref{plant}) by ${\cal C}(A,B)$, i.e.,
$${\cal C}(A,B)=[B,AB,\cdots, A^{n-1}B].$$
It is known {from \cite{R.E1984Between,gu2000new,mengi2008estimation}} that $(A,B)$ is controllable, if and only if ${\cal C}(A,B)$ is of full row rank.

Throughout this paper, given $\bar A \in \{0,*,\times \}^{n\times n}$, $\bar B\in \{0,*,\times \}^{n\times m}$, ${\cal N}_\times$ denotes the set of positions of all
$\times$  entries in $[\bar A, \bar B]$, i.e., ${\cal N}_{\times}=\{(i,j): [\bar A, \bar B]_{ij}=\times\}$. ${\cal N}_{*}=\{(i,j): [\bar A, \bar B]_{ij}=*\}$ is defined similarly. We use $n_*$ and $n_\times$ to denote the numbers of $*$ entries and $\times$ entries in $[\bar A, \bar B]$, respectively. Let $p_1,...,p_{n_*}$ and $\bar p_1,...,\bar p_{n_\times}$ be respectively the parameters for the $*$ entries and $\times$ entries (also called indeterminate parameters) in $[\bar A, \bar B]$. Denote $p=(p_1,...,p_{n_*})$ and $\bar p=(\bar p_1,...,\bar p_{n_{\times}})$. The following notions are natural extensions of the original ones from the real field to the complex field. %{ The following notions are their natural extensions from the real field to the complex field. }   %Moreover, denote $[\tilde A, \tilde B]$ as the generic realization of $[\bar A, \bar B]$, with the parameters in the $(i,j)$th position being $p_{ij}$, $(i,j)\in {\cal N}_{*}\cup {\cal N}_{\times}$. Let $n_*$ and $n_{\times}$ be respectively the numbers of $*$ entries and $\times$ entries in $[\bar A, \bar B]$, i.e., $n_*=|{\cal N}_*|$, $n_\times=|{\cal N}_{\times}|$. , i.e., $n_*=|{\cal N}_*|$, $n_\times=|{\cal N}_{\times}|$
%Now we are introducing the notions of structural controllability and SSC. Depending on the different notions, we shall use different structured matrices to denote the sparsity pattern of $[A,B]$.

\begin{definition}[Structural controllability, {\cite{Lin_1974}}]\label{SC-def} Given \\~ $\bar A \in \{0,*,\times \}^{n\times n}$, $\bar B\in \{0,*,\times \}^{n\times m}$, $(\bar A, \bar B)$ is structurally controllable, if there is an $(A,B)$ satisfying $[A,B]\in {\bf S}_{[\bar A^*, \bar B^*]}$ that is controllable.
\end{definition}

\begin{definition}[SSC, {\cite{strong_controllability}}]\label{SSC-def} Given $\bar A \in \{0,*,\times \}^{n\times n}$, $\bar B \in$\\$\{0,*,\times \}^{n\times m}$, $(\bar A, \bar B)$ is SSC, if $(A,B)$ is controllable for all $[A,B]\in {\bf S}_{[\bar A^\times, \bar B^\times]}$.
\end{definition}

Note that controllability is a generic property in the sense if $(\bar A,\bar B)$ is structurally controllable, where $\bar A\in \{0,*\}^{n\times n}$, $\bar B\in \{0,*\}^{n\times m}$, then $(A,B)$ is controllable for almost all $[A,B]\in {\bf S}_{[\bar A, \bar B]}$. Hereafter, `{\emph{almost all' means `all except for a set of zero Lebesgue measure in the corresponding parameter space'}}.  Motivated by Example \ref{examp-motivate}, as argued in Section \ref{intro-sec}, the PSSC is formally defined as follows.

% The example above motivates a new concept of controllability, namely, PSSC, defined as follows. More precisely, regarding ${\bf S}_{[\bar A,\bar B]}$ as the set of matrices that are parameterized by the parameters $p$, structural controllability of $[\bar A, \bar B]$ means $(A,B)$ is controllable for all $p\in {\mathbb C}^{n_*}\backslash {\cal P}$, where $\cal P$ is some proper variety of the whole parameter space ${\mathbb C}^{n_*}$. %for indeterminate entries of $[\bar A, \bar B]$

%Given $\bar A \in \{0,*,\times\}^{n\times n}$ and $\bar B\in \{0, *,\times\}^{n\times m}$, divide $[\bar A,\bar B]$ into $[\bar A,\bar B]=[\bar A_*,\bar B_*]+[\bar A_\times, \bar B_\times]$, in which `$+$' between two structured matrices is an entry-wise addition operation so that every indeterminate entry in the sum appears exactly once in one of the addends, $\bar A_*\in \{0,*\}^{n\times n}$, $\bar B_*\in \{0,*\}^{n\times m}$, $\bar A_\times \in \{0,\times\}^{n\times n}$, and $\bar B_\times \in \{0,\times\}^{n\times m}$.  With these notations, PSSC is defined as follows.

\begin{definition}[PSSC]\label{PSSC-def} Given $\bar A \in \{0,*,\times\}^{n\times n}$, $\bar B\in$\\$\{0, *,\times\}^{n\times m}$, suppose $[\bar A,\bar B]$ is divided into $[\bar A, \bar B]\!\!=\!\![\bar A_*,\bar B_*]+[\bar A_\times, \bar B_\times]$, in which $\bar A_*\in \{0,*\}^{n\times n}$, $\bar B_*\in \{0,*\}^{n\times m}$, $\bar A_\times \in \{0,\times\}^{n\times n}$, and $\bar B_\times \in \{0,\times\}^{n\times m}$. $(\bar A, \bar B)$ is PSSC, if for almost all $[A_*,B_*]\in {\bf S}_{[\bar A_*,\bar B_*]}$, $(A_*+A_\times,B_*+B_\times)$ is controllable for all $[A_\times, B_\times]\in {\bf S}_{[\bar A_\times, \bar B_\times]}$.
\end{definition}

Equivalently, $(\bar A, \bar B)$ is PSSC, if for all $p\in {\mathbb C}^{n_*}$ except a set of zero measure in ${\mathbb C}^{n_*}$ and all $\bar p\in {\bar {\mathbb C}^{n_\times}}$, the corresponding realization of $(\bar A, \bar B)$ is controllable.  PSSC of $(\bar A, \bar B)$ indicates that, if we randomly generate values for the $*$ entries (from a continuous interval), then with probability $1$, the corresponding system will be controllable for {\emph{all}} nonzero values of the $\times$ entries. For this reason, we may also call an $*$ entry a {\emph{generic entry}}, and the $\times$ one an {\emph{unspecified entry}}. %In Section \ref{sec-genericity}, it is shown that for a single-input system $(\bar A, \bar b)$, if $(\bar A,\bar b)$ is not PSSC, then for almost all $p \in {\mathbb C}^{n_*}$ (or for a generic realization of $p$), there is a $\bar p \in \bar {\mathbb C}^{n_\times}$ so that the corresponding realization is uncontrollable.  Recall $p$ and $\bar p$ are respectively the parameters for the $*$ entries and $\times$ entries in $[\bar A, \bar B]$.

To show PSSC is well-defined, we introduce the concept \\~ perturbation-tolerant structural controllability (PTSC) from \cite{full-version-tac}, which is defined for a structured system as the property that, for almost all values of its generic entries, there exist no complex values ({\emph{including zero}) for its unspecific entries such that the corresponding system realization is uncontrollable (see Definition 3 of \cite{full-version-tac} for a precise description). It is proven in \cite{full-version-tac} that, depending solely on the structure of the structured system, for almost all values of its generic entries, either there exist no values for its unspecific entries that can make the system realization uncontrollable, or there exist such values. Obviously, PTSC is a sufficient condition for the property in Definition \ref{PSSC-def} to hold. This means, there do exist structured systems satisfying the property in Definition \ref{PSSC-def}.

%In other words, $(\bar A, \bar B)$ is PSSC, if for almost all values of the $*$ entries and all nonzero values for the $\times$ entries, the corresponding system is controllable. That is to say,

Further, it is easy to see, if no $*$ entries exist in $(\bar A, \bar B)$, then PSSC collapses to SSC, which explains the term `partial' in this terminology.  On the other hand, if no $\times$ entries exist in $(\bar A, \bar B)$, then PSSC collapses to structural controllability (this also demonstrates that PSSC is well-defined). In this sense, PSSC bridges the gap between structural controllability and SSC. Moreover, it is apparent that for a given $(\bar A, \bar B)$, SSC implies PSSC, and PSSC implies structural controllability\footnote{Note as controllability is a generic property \cite{generic}, if a structured system is structurally controllable, there must exist a controllable realization of it subject to the constraint that an arbitrary subset of its indeterminate entries are nonzero.}; while the inverse direction is not necessarily true. The following Example \ref{SSC-PSSC} highlights the differences among these concepts.

\begin{example}\label{SSC-PSSC} Consider a single-input structured system as {\small
\begin{equation}\label{example-sys1}\bar A=\left[
  \begin{array}{cccc}
    0 & 0 & 0 & 0 \\
    * & 0 & 0 & 0 \\
    0 & \times & 0 & 0 \\
    * & * & 0 & * \\
  \end{array}
\right], \bar b=\left[
                  \begin{array}{c}
                    \times \\
                    0 \\
                    0 \\
                    0 \\
                  \end{array}
                \right].
 \end{equation}}Let $[\tilde A, \tilde b]$ be a generic realization of $[\bar A, \bar b]$ by assigning $a_{ij}$ (resp. $b_{ij}$) to the $(i,j)$th indeterminate entry of $\bar A$ (resp. $\bar b$). We have
\begin{equation}\label{det-expression} \det {\cal C}(\tilde A, \tilde b)= a^2_{21}a_{32}b_{11}^4a_{44}(a_{41}a_{44}+a_{21}a_{42}). \end{equation} Hence, $(\bar A, \bar b)$ is structurally controllable.   Moreover, in case \\~ $a_{41}a_{44}+a_{21}a_{42}=0$, the obtained system will be uncontrollable, indicating that $(\bar A, \bar b)$ is not SSC. However, for all $(a_{21}, a_{41}, a_{42}, a_{44})\in {\mathbb C}^4$ except $\{(a_{21}, a_{41}, a_{42}, a_{44})\in {\mathbb C}^4: a_{21}^2a_{44}(a_{41}a_{44}+a_{21}a_{42})= 0\}$, whatever nonzero values $a_{32}$ and $b_{11}$ take, the obtained system is controllable. This means $(\bar A, \bar b)$ is PSSC. \hfill $\square$
 \end{example}
%The following example highlights the difference between PSSC and SSC.

 %This means, $(\bar A, \bar b)$ is PSSC.

\begin{remark}[Complex field and real field] It is noted that in Definition \ref{PSSC-def}, PSSC is defined for $(A,B)$ in the complex-filed. As shown in Section \ref{sec-genericity}, this can guarantee that if $(\bar A,\bar b)$ is not PSSC, then for almost all values of its $*$ entries, there exist nonzero values for its $\times$ entries, such that the corresponding realization is uncontrollable. We may also define PSSC in the real field, by restricting that all of the indeterminate parameters should take real values. { However, as explained in Remark \ref{remark-2}, it then may happen that even for single-input systems, this property will not solely depend on the combinatorial properties of system structures. By Definition \ref{PSSC-def} and from the property of algebraic independence \cite{dummit2004abstract}, it is not hard to see, PSSC in the complex field is sufficient for the same property to hold in the real field.}% {We shall further explain the motivation of defining PSSC in the complex field in Remark \ref{remark-2}.} % {See Remark \ref{remark-2} for further explanations for the definition of PSSC in the complex field.}
\end{remark}

\subsection{Preliminaries } \label{sub-preli}
%\subsubsection{Graph theory}
%The following will introduce some preliminaries in graph and structured system theories.

A (directed) graph ${\cal G}$ is represented by ${\cal G}=(V,E)$, where $V$ is the vertex set and $E\subseteq V\times V$ is the edge set. If there is a path (i.e., a sequence of successive edges) from vertex $v_j$ to vertex $v_i$, we say $v_i$ is reachable from $v_j$. A subgraph ${\cal G}_s=(V_s,E_s)$ of $\cal G$ is a graph such that $V_s\subseteq V$ and $E_s\subseteq E$. And ${\cal G}_s$ is said to be induced by $V_s$ if $E_s=(V_s\times V_s)\cap E$. For $V_s\subseteq V$, {${\cal G}-V_s$ denotes the graph obtained from ${\cal G}$ by deleting the vertices in $V_s$ together with their incident edges (i.e., ${\cal G}-V_s$ is the subgraph of ${\cal G}$ induced by $V\backslash V_s$)}.

% If $V$ of a graph $\cal G$ can be partitioned into two subsets (called bipartitions), such that there is no edge of $E$ connecting the vertices within each subset, then $\cal G$ is called bipartite.

A bipartite graph is written as ${\cal G}=(V^+,V^-,E)$ with $V^+$ and $V^-$ being the bipartitions and $E$ the edges \cite{Murota_Book}. For a $v\in V^+\cup V^-$, ${\cal N}({\cal G},v)$ denotes the set of neighbors in $\cal G$, i.e., the set of vertices that are connected with $v$ by an edge.  A {\emph{matching}} of ${\cal G}$ is a set of its edges, no two of which share a common end vertex. The size (or cardinality) of a matching is the number of edges it contains. A vertex is said to be {\emph{matched}} by a matching if it is an end vertex of edges in this matching.  A {\emph{maximum matching}} of $\cal G$ is the matching with the largest size, which value is denoted by ${\rm {mt}}({\cal G})$. An edge of $\cal G$ is said to be {\emph{admissible}} if it is contained in some maximum matching of ${\cal G}$. We say a matching ${\cal M}$ {\emph{covers}} ${\cal G}$, if ${\cal M}\cap E$ forms a maximum matching of ${\cal G}$. If each edge of $\cal G$ has a non-negative weight, the {\emph{maximum (minimum) weight maximum matching}} is the maximum matching of $\cal G$ with the largest (smallest) weight (the weight of a matching is the sum of its edge weights).

%Dulmage-Mendelsohn decomposition (DM-decomposition) is to decompose a bipartite graph into subgraphs with respect to maximum matchings. %For our derivations, only a simplified version of DM-decomposition, i.e., DM-decomposition of bipartite graphs with a perfect matching,  is needed; see the following definition.

\begin{definition}[DM-decomposition, \cite{Murota_Book}] \label{DM-def} Given a bipartite graph ${\cal G}=(V^+,V^-,E)$, the Dulmage-Mendelsohn decomposition (DM-decomposition) of ${\cal G}$ is to decompose ${\cal G}$ into subgraphs ${\cal G}_i=(V_i^+,V_i^-,E_i)$ ($i=0,1,...,d,\infty$, each ${\cal G}_i$ is called a DM-component) satisfying:

1) $V^\star=\bigcup \nolimits_{i=0}^\infty V_i^\star$, $V_i^\star\bigcap V_j^\star=\emptyset$ for $i\ne j$, with $\star=+,-$; $E_i=\{(v^+,v^-)\in E: v^+\in V_i^+, v^-\in V_i^-\}$;

2) ${\rm mt}({\cal G}_i)=|V_i^+|=|V_i^-|$ for $i=1,...,d$, ${\rm mt}({\cal G}_0)=|V^+_0|<|V^-_{0}|$ if $V^+_0\ne \emptyset$, and ${\rm mt}({\cal G}_\infty)=|V^-_{\infty}|< |V^+_\infty|$ if $V^-_\infty\ne \emptyset$; Moreover, each $e\in E_i$ is admissible in ${\cal G}_i$, $i=0,1,...,d,\infty$.

3) $E_{ij}=\emptyset$ unless $0\le i \le j \le \infty$, and $E_{ij}\ne \emptyset$ only if $0\le i \le j \le \infty$, where $E_{ij}=\{(v^+,v^-)\in E: v^+\in V^+_i, v^-\in V^-_j\}$;

4) ${\cal M}$ ($\subseteq {\cal E}$) is a maximum matching of $\cal G$ iif ${\cal M}\subseteq \bigcup \nolimits_{i=0}^{\infty} {\cal E}_k$ and ${\cal M}\cap {\cal E}_i$ is a maximum matching of ${\cal G}_i$ for $i=0,...,\infty$;

5) ${\cal G}$ cannot be decomposed into more components satisfying conditions 1)-4).
\end{definition}

In Definition \ref{DM-def}, ${\cal G}_0$ (if exists) is called the horizontal tail, and ${\cal G}_\infty$ the vertical tail. Additionally, if ${\cal G}$ contains only one DM-component, then ${\cal G}$ is called DM-irreducible.

For an $n_1\times n_2$ matrix $M$, the bipartite graph associated with $M$ is given by ${\cal B}(M)=(V^+,V^-,E_M)$, where $V^+=\{v^+_1,...,v^+_{n_1}\}$ (resp. $V^-=\{v^-_1,...,v^-_{n_2}\}$) corresponds to the rows (columns) of $M$, and $E_M=\{(v^+_i,v^-_j): M_{ij}\ne 0, v^+_i\in V^+, v^-_j\in V^-\}$. It is known that, for a generic matrix $M$, ${\rm grank}(M)={\rm mt}({\cal B}(M))$ \cite{Murota_Book}.  For $V_s^+\subseteq V^+$ and $V_s^-\subseteq V^-$, we also write $M[V_s^+,V_s^-]$ as the submatrix of $M$ whose rows correspond to $V_s^+$ and columns to $V_s^-$, where elements in $V_s^+$ and $V_s^-$ are ordered. Note such an expression of submatrices is {\emph{invariant}} subject to row and column permutations on $M$, that is, upon letting $P\in {\mathbb R}^{n_1\times n_1}$ and $Q\in {\mathbb R}^{n_2\times n_2}$ be two permutation matrices and $M'=PMQ$, $M'[V^+_s,V^-_s]$ is the same as $M[V^+_s,V^-_s]$.

%Recall that a multivariable polynomial $f$ is irreducible if it cannot be factored as $f=f_1f_2$ where $f_1,f_2$ have degrees smaller than $f$.
%\begin{lemma} Let ${\cal B}(M)$ be a bipartite graph associated with a generic square matrix $M$. Then, $\det M$ is irreducible, if and only if ${\cal B}(M)$ is DM-irreducible. (noting $x_i$ for $i=n+1,...,n+m$ is itself input-reachable)
%\end{lemma}

%\subsubsection{Structured system theory}
For a structured {pair} $(\bar A, \bar B)$, its associated graph is given by ${\cal G}(\bar A, \bar B)=(X, E_{X})$, where $X=\{x_1,...,x_n,...,x_{n+m}\}$, $E_X=\{(x_j,x_i): [\bar A,\bar B]_{ij}\ne 0\}$. A vertex ${ x_i\in X}$ is said to be input-reachable, if ${x_i}$ is reachable from at least one vertices of $x_{n+1},...,$ $x_{n+m}$ in ${\cal G}(\bar A, \bar B)$.
\begin{lemma}\cite{generic} \label{structural-controllability-condition} Given $\bar A\in \{0,*\}^{n\times n}$, $\bar B\in \{0,*\}^{n\times m}$, $(\bar A, \bar B)$ is structurally controllable, if and only if i) every vertex $x\in X$ is input-reachable, and ii) ${\rm grank}[\bar A, \bar B]=n$.
\end{lemma}

\section{Properties and conditions of PSSC for single-input systems} \label{single-case} % Main Results
%The first one is in terms of
In this section, we first present some properties of PSSC of single-input systems. Particularly, we show PSSC generalizes the generic property embedded in the conventional structural controllability. We then give a computationally efficient graph-theoretic criterion for PSSC.

\subsection{Properties} \label{sec-genericity}
We first give an algebraic condition for PSSC, in terms of determinants of the controllability matrix of the generic realization of $(\bar A, \bar b)$.
%either for almost all $[A_*,b_*]\in {\bf S}_{[\bar A_*,\bar b_*]}$, $(A_*+A_\times,b_*+b_\times)$ is controllable for every $[A_\times,b_\times]\in {\bf S}_{[\bar A_\times, \bar b_{\times}]}$, or for almost all $[A_*,b_*]\in {\bf S}_{[\bar A_*, \bar b_*]}$, there is a $[A_\times,b_\times]\in {\bf S}_{[\bar A_\times, \bar b_{\times}]}$ such that $(A_*+A_\times, b_*+b_\times)$ is uncontrollable. %This makes sense of the notion of PSSC in Definition \ref{PSSC-def}.

%Let $n_*$ and $n_\times$ be respectively the number of $*$ and $\times$ entries in $[\bar A, \bar b]$. Moreover, let $[\tilde A, \tilde b]$ be a generic realization of $[\bar A, \bar b]$, with the parameters for the $*$ entries being $p_1,...,p_{n_*}$ and the parameters for the $\times$ entries being $\bar p_1,...,\bar p_{n_\times}$. Denote $p=(p_1,...,p_{n_*},\bar p_1,...,\bar p_{n_\times})$. %Recall that a monomial in a polynomial is a product of powers of variables with nonnegative integer exponents (degrees).

%Let $[\tilde A, \tilde b]$ be a generic realization of $[\bar A, \bar b]$, with the parameters for the $*$ entries being $p_1,...,p_{n_*}$ and for the $\times$ entries being $\bar p_1,...,\bar p_{n_\times}$, where $n_*$ and $n_\times$ are the numbers of $*$ and $\times$ entries in $[\bar A, \bar b]$, respectively.  $(\bar A, \bar b)$ is PSSC, if and only if $\det {\cal C}(\tilde A, \tilde b)\ne 0$ and has the following form
%$$\det {\cal C}(\tilde A, \tilde b)=f(p_1,...,p_{n_*})\prod \nolimits_{i=1}^{n_\times} \bar p_i^{r_i}, $$
%where $f(p_1,...,p_{n_*})$ is a polynomial of $p_1,...,p_{n_*}$, $r_i\ge 0$.

\begin{theorem}\label{algebraic-condition}Given $\bar A \in \{0,*,\times\}^{n\times n}$, $\bar b\in \{0, *,\times\}^{n\times 1}$, let $[\tilde A, \tilde b]$ be a generic realization of $[\bar A, \bar b]$, with the parameters for the $*$ entries being $p_1,...,p_{n_*}$ and for the $\times$ entries being $\bar p_1,...,\bar p_{n_\times}$. $(\bar A, \bar b)$ is PSSC, if and only if $\det {\cal C}(\tilde A, \tilde b)\ne 0$ and has the following form
\begin{equation}\label{formal-form} \det {\cal C}(\tilde A, \tilde b)=f(p_1,...,p_{n_*})\prod \nolimits_{i=1}^{n_\times} \bar p_i^{r_i}, \end{equation}
where $f(p_1,...,p_{n_*})$ is a nonzero polynomial (including the constant $1$) of $p_1,...,p_{n_*}$, $r_i\ge 0$.
\end{theorem}

Although the verification of Theorem \ref{algebraic-condition} is prohibitive for large-scale systems, as computing the determinant of a symbolic matrix has computational complexity increasing exponentially with its dimensions \cite{A_Shpilka_survey}, it is theoretically significant in proving some properties of PSSC.  Particularly, based on Theorem \ref{algebraic-condition}, we have the following two properties of PSSC of single-input systems.

\begin{proposition}\label{pro-generic} Given $\bar A\in \{0,*,\times\}^{n\times n}$, $\bar b\in \{0,*,\times\}^{n\times 1}$, let $[\bar A_*,\bar b_*]$ and $[\bar A_\times, \bar b_\times]$ be defined in Definition \ref{PSSC-def}. Then, depending on $[\bar A, \bar b]$, either for almost all $[A_*,b_*]\in {\bf S}_{[\bar A_*,\bar b_*]}$, $(A_*+A_\times,b_*+b_\times)$ is controllable for every $[A_\times,b_\times]\in {\bf S}_{[\bar A_\times, \bar b_{\times}]}$, or for almost all $[A_*,b_*]\in {\bf S}_{[\bar A_*, \bar b_*]}$, there is a $[A_\times,b_\times]\in {\bf S}_{[\bar A_\times, \bar b_{\times}]}$ such that $(A_*+A_\times, b_*+b_\times)$ is uncontrollable.
\end{proposition}

The above proposition reveals for a single-input structured system, if it is not PSSC, then for almost all values of the $*$ entries, there exist nonzero (complex) values for the $\times$ entries so that the realization is uncontrollable; otherwise, for almost all values of the $*$ entries, the corresponding realization is controllable for all nonzero values of the $\times$ entries. This generalizes the generic property embedded in structural controllability. {This proposition also explains the motivation behind the definition of PSSC.} %for almost all values of the $*$ entries, the corresponding realization is controllable for all nonzero values of the $\times$ entries; otherwise,

%PSSC can characterize the generic property that
%{\bf please check}
\begin{remark}[PSSC in real field] \label{remark-2} From Theorem \ref{algebraic-condition}, it is easy to see, if $(\bar A,\bar b)$ is PSSC, then for {\emph{almost all}} $p\in {\mathbb R}^{n_*}$ and {\emph{all}} $\bar p\in \bar {\mathbb R}^{n_\times}$, the corresponding realization is controllable. However, if $(\bar A,\bar b)$ is not PSSC, it is possible that for $p$ in a full-dimensional semi-algebraic set ${\cal P}^*$ of ${\mathbb R}^{n_*}$ (i.e., a subset of ${\mathbb R}^{n_*}$ defined by a finite sequence of polynomial equations and inequalities, or their finite unions \cite{dummit2004abstract}), the corresponding realization of $(\bar A,\bar b)$ is controllable for all $\bar p\in \bar {\mathbb R}^{n_\times}$; while for $p$ in the semi-algebraic set ${\mathbb R}^{n_*}\backslash {\cal P}^*$ of full dimension, there is a $\bar p\in \bar {\mathbb R}^{n_\times}$ that makes the corresponding realization uncontrollable. This is because, as argued in the proof of Theorem \ref{algebraic-condition}, $\det {\cal C}(\tilde A, \tilde b)$ then has a form of (\ref{non-PSSC-form}), which does not necessarily have real solutions for $\bar p_i$.  { Therefore, {\emph{the study of PSSC in the complex filed is not only of theoretical significance itself, as it characterizes how certain combinatorial properties of the system structure affect the robustness of controllability (at least for single-input systems), but also of practical significance for the robustness evaluation of controllability in the real filed, particularly when only the system structure is available but the semi-algebraic subset of its parameters.}}} %{ In addition, if $(\bar A, \bar b)$ is not PSSC, for any proper algebraic variety ${\cal P}$ of ${\mathbb R}^{n_*}$, there exists at least one $p\in {\mathbb R}^{n_*}\backslash {\cal P}$ and $\bar p_i\in {\mathbb R}\backslash \{0\}$ ($i=1,...,n_\times$), so that the corresponding realization of $(\bar A, \bar b)$ is uncontrollable. }
\end{remark}

%Based on Theorem \ref{algebraic-condition}, the following property is crucial to the derivations of conditions for PSSC, which transforms the problem of verifying PSSC of $(\bar A, \bar b)$ to $n_\times$ subproblems of verifying PSSC of

\begin{proposition} \label{one-edge-principle} Let $(\bar A^\pi, \bar b^\pi)$ be the structured system obtained from $(\bar A,\bar b)$ by preserving its $(i,j)$th $\times$ entry and replacing the remaining indeterminate entries with $*$ entries, for $\pi\doteq (i,j)\in {\cal N}_\times$, i.e., $[\bar A^\pi, \bar b^\pi]_{lk}=*$ if $[\bar A, \bar b]_{lk}\ne 0$ and $(l,k)\ne \pi$, and $[\bar A^\pi, \bar b^\pi]_{lk}=[\bar A, \bar b]_{lk}$ otherwise. Then, $(\bar A, \bar b)$ is PSSC, if and only if for each $\pi \in {\cal N}_\times$, $(\bar A^\pi,\bar b^\pi)$ is PSSC.
\end{proposition}

The above proposition is beneficial in deriving conditions for PSSC, as it actually transforms the problem of verifying PSSC of $(\bar A, \bar b)$ to $n_\times$ subproblems of verifying PSSC of $(\bar A^\pi,\bar b^\pi)$ for each $\pi\in {\cal N}_\times$.
 %Currently, we are not able to extend this proposition to the multi-input case ()

%Conditions of PSSC for Single-Input Systems}
\subsection{Necessary and sufficient conditions}
In this subsection, we give testable necessary and sufficient conditions for single-input systems to be PSSC. Owing to Proposition \ref{one-edge-principle}, in the following, we first focus on conditions of PSSC for $(\bar A^\pi,\bar b^\pi)$, i.e. systems that contain only one $\times$ entry, and then on the general systems.
Recall from the PBH test \cite{R.E1984Between,mengi2008estimation}, an uncontrollable mode for system $(A,B)$ is a $\lambda \in {\mathbb C}$ such that ${\rm rank}([A-\lambda I, B])<n$. Inspired by \cite{strong_controllability,full-version-tac}, we shall respectively give the conditions for the nonexistence of zero uncontrollable and nonzero uncontrollable modes. Particularly, the following lemma, relating the `structure' of a vector in the left null space of a given matrix to the full row rank of its submatrices, is fundamental to our subsequent derivations. %sparsity pattern

\begin{lemma}[Lemma 7 of \cite{zhang2021generic}]\label{null-space} Given $M\in {\mathbb C}^{n_1\times n_2}$, let $T$ consist of $r$ {\emph{linearly independent row vectors}} that span the left null space of $M$. Then, for any $K\subseteq J_{n_1}$, $T[J_r,K]$ is of full row rank, if and only if $M[J_{n_1}\backslash K, J_{n_2}]$ is of full row rank.
\end{lemma}% (i.e., $T$ is a basis matrix of the left null space of $M$)

\begin{proposition}\label{zero-condition} Suppose $(\bar A,\bar b)$ is structurally controllable and $[\bar A, \bar b]$ contains only one $\times$ entry, with its position being $(i,j)$. Moreover, let $[\bar A, \bar b]$ be divided into $[\bar A_*, \bar b_*]+[\bar A_\times, \bar b_\times]$ in the way described in Definition \ref{PSSC-def}.
For almost all $[A_*,b_*]\in {\bf S}_{[\bar A_*,\bar b_*]}$, there exist no $[A_\times, b_\times]\in {\bf S}_{[\bar A_\times, \bar b_\times]}$ and nonzero vector $q\in {\mathbb C}^n$ that satisfy $q^\intercal [A_*+A_\times, b_*+b_\times]=0$, if and only if one of the following conditions holds:

a1) ${\rm grank}([\bar A, \bar b][J_n,J_{n+1}\backslash \{j\}])=n$;

a2) For each $k\in {\cal N}_*^j$, ${\rm grank}([\bar A, \bar b][J_n\backslash \{k\},J_{n+1}\backslash \{j\}])=n-2$, in which ${\cal N}_{*}^j=\{k\in J_n\backslash \{i\}: [\bar A, \bar b]_{kj}= *\}$;\footnote{If ${\cal N}_*^j=\emptyset$, this condition is set to be satisfied (the same below).} %${\rm grank}([\bar A_*, \bar b_*][J_n\backslash \{i\},J_{n+1}\backslash \{j\}])=n-1$ and %${\rm grank}([\bar A_*, \bar b_*][J_n\backslash \{i\},J_{n+1}\backslash \{j\}])=n-1$, and for

a3) ${\rm grank}([\bar A, \bar b][J_n\backslash \{i\},J_{n+1}\backslash \{j\}])=n-2$.
\end{proposition}

Proposition \ref{zero-condition} gives an algebraic condition for the nonexistence of zero uncontrollable modes. % Since deleting a column from a matrix can reduce its rank by at most one, condition a1) and condition a2) with ${\cal N}_*^j\ne \emptyset$ (or condition a3)) cannot be fulfilled simultaneously. Condition a2) and condition a3), on the other hand, can hold simultaneously.
The following proposition gives an equivalent bipartite graph form of Proposition \ref{zero-condition}. Recall the bipartite graph ${\cal B}([\bar A, \bar b])=(V^+,V^-,E_{[\bar A, \bar b]})$ associated with $[\bar A, \bar b]$ is defined in Section \ref{sub-preli},  where $V^+=\{v_1^+,...,v_n^+\}$, $V^-=\{v_1^-,...,v_{n+1}^-\}$. %Recall ${\cal B}([\bar A, \bar b])=(V^+,V ^-,E_{[\bar A, \bar b]})$, where $V^+=\{v_1^+,...,v_n^+\}$, $V^-=\{v_1^-,...,v_{n+1}^-\}$, and $E_{[\bar A, \bar b]}=\{(v^+_k,v^-_l): [\bar A,\bar b]_{kl}\ne 0\}$.% for the structured matrix $[\bar A, \bar b]$

\begin{proposition} \label{graph-zero-mode}
Under the same setting in Proposition \ref{zero-condition}, conditions a1)-a3) are respectively equivalent to

b1) ${\cal B}([\bar A, \bar b])$ contains a maximum matching that does not match $v_j^-$;

b2) Either ${\cal N}({\cal B}([\bar A, \bar b]),v_j^-)=\{v_i^+\}$ or for each $v_k^+\in$\\~${\cal N}({\cal B}([\bar A, \bar b]),v_j^-)\backslash \{v_i^+\}$, ${\rm mt}({\cal B}([\bar A, \bar b])-\{v_k^+,v_j^-\})=n-2$;

b3) ${\rm mt}({\cal B}([\bar A, \bar b])-\{v_i^+,v_j^-\})=n-2$.
\end{proposition}

% ${\rm grank}([\bar A, \bar b][J_n\backslash \{i\},J_{n+1}\backslash \{j\}])=n-2$.

% Among $l\in \{l: [\bar A,\bar b]_{lj}\ne 0\}$, there is at most one vertex, being exactly $v_i^+$, that satisfies ${\rm grank}([\bar A, \bar b][J_n\backslash \{i\},J_{n+1}\backslash \{j\}])=n-1$;

%For conditions of nonzero uncontrollable modes, we rely on the DM-decomposition.
%The DM-decomposition of ${\cal B}(M)=(V^+,V^-,E)$ into $G_i=(V_i^+,V_i^-,E_i)$ ($i=1,...,d$), supposing ${\cal B}(M)$ has a perfect matching, corresponds to that the matrix $M$ is transformed into the block-triangular form via two permutation matrices $P$ and $Q$:
%$$PMQ=\left[
%        \begin{array}{ccc}
%          M_{1} & \cdots & M_{1d} \\
%          0 & \ddots & \vdots \\
%          0 & 0 & M_d \\
%        \end{array}
%      \right],
%$$where the submatrix $M_i=M[V^+_i,V^-_i]$ corresponds to $G_i$ ($i=1,...,d$).

We are now presenting the conditions for nonzero uncontrollable modes. Again, suppose $(\bar A,\bar b)$ is structurally controllable and $[\bar A, \bar b]$ contains only one $\times$ entry, with its position being $(i,j)$. For a generic realization $(\tilde A, \tilde b)$ of $(\bar A, \bar b)$, define a generic matrix pencil $H_\lambda=[\tilde A-\lambda I, \tilde b]$, and let $H^{j_c}_\lambda\doteq H_\lambda[J_n,J_{n+1}\backslash \{j\}]$ and $H^j_\lambda\doteq H_\lambda[J_n,\{j\}]$. Let ${\cal B}(H_\lambda)=(V^+,V^-,E_{H_\lambda})$ be the bipartite graph associated with $H_{\lambda}$, defined as follows: $V^+=\{v^+_1,...,v^+_n\}$, $V^-=\{v^-_1,...,v^-_{n+1}\}$,
 and $E_{H_\lambda}=E_I\cup E_{[\bar A, \bar b]}$, in which $E_I=\{(v^+_k,v^-_k): k=1,...,n\}$ and $E_{[\bar A,\bar b]}=\{(v^+_k,v^-_l): [\bar A,\bar b]_{kl}\ne 0\}$. An edge $e$ is called a {\emph{$\lambda$-edge}} if $e\in {E}_I$, and a {\emph{self-loop}} if $e\in E_I\cap E_{[\bar A, \bar b]}$. ${\cal B}(H_\lambda^{j_c})$ is the subgraph of ${\cal B}(H_\lambda)$ induced by $V^+\cup V^-\backslash \{v^-_j\}$.

Let ${\cal G}^{j_c}_k=(V^+_k,V^-_k,E_k)$ ($k=0,...,d,\infty$) be the DM-components of ${\cal B}(H_\lambda^{j_c})$. From \citep[Lem 4]{full-version-tac}, as $(\bar A, \bar b)$ is structurally controllable, we have ${\rm mt}({\cal B}(H_\lambda^{j_c}))=n$ for $j=1,...,n+1$. Hence, ${\cal G}^{j_c}_0={\cal G}^{j_c}_\infty=\emptyset$, $\forall j\in J_{n+1}$.  %^Moreover, for each $l\in \{1,...,n\}$, suppose vertex $v^+_l$ corresponds to the
By the correspondence between a matrix and its associated bipartite graph, the DM-decomposition of ${\cal B}(H_\lambda^{j_c})$ corresponds to that the matrix $H_\lambda^{j_c}$ is transformed into the block-triangular form via two $n\times n$ permutation matrices $P$ and $Q$ \cite{Murota_Book}:
\begin{equation}\label{DM-decomp} PH_\lambda^{j_c}Q=\left[
        \begin{array}{ccc}
          M^{j_c}_{1}(\lambda) & \cdots & M^{j_c}_{1d}(\lambda) \\
          0 & \ddots & \vdots \\
          0 & 0 & M^{j_c}_d(\lambda) \\
        \end{array}
      \right]\doteq M_\lambda^{j_c},
\end{equation} where the submatrix $M^{j_c}_{k}(\lambda)=H_\lambda^{j_c}[V^+_k,V^-_k]$ corresponds to ${\cal G}_k$ ($k=1,...,d$). Accordingly, let $M_\lambda^j\doteq PH_\lambda^j$.
% (hereafter, for notation simplicity, we have used $H_\lambda^{j_c}[V^+_k,V^-_k]$ the denote the submatrix of $H_\lambda^{j_c}$ in the rows corresponding to $V^+_k$ and columns corresponding to $V^-_k$)

For each $k\in \{1,...,d\}$, let $\gamma_{\min}({\cal G}^{j_c}_k)$ and $\gamma_{\max}({\cal G}^{j_c}_k)$ be respectively the minimum and maximum numbers of $\lambda$-edges contained in a matching over all maximum matchings of ${\cal G}^{j_c}_k$. We borrow the boolean function $\gamma_{nz}(\cdot)$ for ${\cal G}_k^{j_c}$ from \cite{full-version-tac}, which is defined as
\begin{equation}\label{fun-nz} \gamma_{\rm nz}({\cal G}_k^{j_{\rm c}})=
\begin{cases} 1 & {\begin{array}{c} \text{if}\ \gamma_{\max}({\cal G}_k^{j_{\rm c}})-\gamma_{\min}({\cal G}_k^{j_{\rm c}})>0 \\ \text{or} \ {\cal G}_k^{j_{\rm c}} \ {\text{contains a self-loop}} \end{array}} \\
0 &  \text{otherwise}.
\end{cases}\end{equation}
The following lemma explains the motivation of introducing $\gamma_{\rm nz}(\cdot)$.

\begin{lemma}[Lemma 9 of \cite{full-version-tac}]\label{nonzero-function}  With notations as above, \\~$\det M_k^{j_c}(\lambda)$ ($k\in \{1,...,d\}$) generically has nonzero roots for $\lambda$, if and only if $ \gamma_{\rm nz}({\cal G}_k^{j_{\rm c}})=1$.
\end{lemma}

%It has been proved in \cite{full-version-tac} that, $\det $

Next, define the set
\begin{equation}\label{important-set} {\Omega}_j=\{k\in \{1,...,d\}: {\gamma}_{\rm nz}({\cal G}^{j_{\rm c}}_k)= 1\}.\end{equation}
Moreover, for a vertex $v^+_l\in V^+$,  define a set $\Omega^l_{j}\subseteq \Omega_j$ as
\begin{equation}\label{important-subset} \begin{array}{c}  \Omega^l_{j}=\left\{k\in \Omega_j: {\cal G}_k^{j_c} \ {\text{ is not covered by at least}} \right.\\
 \left.{\text{one maximum matching of }}  \ {\cal B}(H_\lambda^{j_c})-\{v^+_l\}  \right\}  \end{array}.\end{equation}
It is clear $\Omega_j^l=\emptyset$ implies that every maximum matching of ${\cal B}(H_\lambda^{j_c})-\{v^+_l\}$ covers $\bigcup \nolimits_{k\in \Omega_j} {\cal G}_k^{j_c}$. {As we shall see, $\Omega_j^l$ is the set of indices of all DM-components associated with which $\det M_k^{j_c}(\lambda)$ has a nonzero root $z$, such that $M_z^{j_c}[V^+\backslash \{v_l^+\},V^-]$ is of full row rank ($M_z^{j_c}$ is obtained by substituting $\lambda=z$ into $M_\lambda^{j_c}$). Hereafter, by saying $M_{\lambda}^{j_c}$ (or its sub-matrices) satisfies certain properties, we mean these properties are satisfied for almost all values of the corresponding indeterminate parameters (i.e., they are satisfied generically).}

%The following lemma involving $\Omega_j^l$ is crucial to characterizing the nonzero uncontrollable modes. In this lemma, by saying $M_{\lambda}^{j_c}$ (or its sub-matrices) satisfies certain properties, we mean these properties are satisfied for almost all values of the corresponding indeterminate parameters (i.e., they are satisfied generically). %It is clear $\Omega_j=\emptyset$ implies $\Omega_j^l=\emptyset$.% In the special case $\Omega_j=\emptyset$, we have $\Omega_j^l=\emptyset$.

%

\begin{lemma}\label{row-rank-deficient} Let $M_\lambda^{j_c}$ be defined in (\ref{DM-decomp}). Assume $\Omega_j\ne \emptyset$.
The following properties are true:

1) Given a $v_l^+\in V^+$, $M_{z}^{j_c}[V^+\backslash \{v^+_l\}, V^-]$ is generically row rank deficient for {\bf all} $z\in \{\lambda\in {\mathbb C}\backslash \{0\}:\det M_\lambda^{j_c}=0\}$, if and only if $\Omega_j^l=\emptyset$;\footnote{Please note, since $M_z^{j_c}$ does not contain the column corresponding to $v_j^-$,  $M_z^{j_c}[V^+\backslash \{v_l^+\}, V^-]$ is the same as $M_z^{j_c}[V^+\backslash \{v_l^+\}, V^-\backslash \{v_j^-\}]$. The same case holds for other matrices.} %(the same for $H_z^{j_c}$)

2) Suppose $\Omega_j^i \ne \emptyset$. For all nonzero $z$ making \\ $M_{z}^{j_c}[V^+\backslash \{v^+_i\}, V^-]$ of full row rank and $\det M_z^{j_c}=0$ simultaneously,  $M_{z}^{j_c}[V^+\backslash \{v^+_{l}\}, V^-]$ is not of full row rank for a given $v_l^+\in V^+\backslash \{v_i^+\}$, if and only if every maximum matching of $B(H_\lambda^{j_c})-\{v_l^+\}$ covers $\bigcup \nolimits_{k\in \Omega_j^i} {\cal G}_k^{j_c}$.
%there exists $z\in {\mathbb C}\backslash \{0\}$ making $M_{z}^{j_c}[V^+\backslash \{v^+_{l_1}\}, V^-]$ and $M_{z}^{j_c}[V^+\backslash \{v^+_{l_2}\}, V^-]$ generically of full row rank simultaneously, while $\det(M_z^{j_c})=0$.
\end{lemma}

%\begin{definition}
%A vertex $v^+_l\in V^+$ is said to satisfy condition ({\bf$\star$}), if every maximum matching of ${\cal B}(H_\lambda^{j_c})-\{v^+_l\}$ covers $\bigcup \nolimits_{k\in \Omega_j} G_k^{j_c}$. In the special case $\Omega_j=\emptyset$, every vertex $v^+_l\in V^+$ satisfies condition ({\bf$\star$}).
%\end{definition}
%
%If vertex $v^+_l\in V^+$ does not satisfy condition ({\bf$\star$}), define a set $\Omega^l_{j}\subseteq \Omega_j$ as
%$$\begin{array}{c}  \Omega^l_{j}=\left\{k\in \Omega_j: G_k^{j_c} \ {\text{ is not covered by at least}} \right.\\
% \left.{\text{one maximum matching of }}  \ {\cal B}(H_\lambda^{j_c})-\{v^+_l\}  \right\}  \end{array}.$$

%From the block-triangular structure of $M_\lambda^{j_c}$,

\begin{proposition}\label{nonzero-condition} Suppose $(\bar A,\bar b)$ is structurally controllable and $[\bar A, \bar b]$ contains only one $\times$ entry, with its position being $(i,j)$. Let $[\bar A, \bar b]$ be divided into $[\bar A_*, \bar b_*]+[\bar A_\times, \bar b_\times]$ in the way described in Definition \ref{PSSC-def}.
For almost all $[A_*,b_*]\in {\bf S}_{[\bar A_*,\bar b_*]}$, there exist no $[A_\times, b_\times]\in {\bf S}_{[\bar A_\times, \bar b_\times]}$, nonzero complex number $z$ and nonzero vector $q\in {\mathbb C}^n$ that satisfy $q^\intercal [A_*+A_\times-z I, b_*+b_\times]=0$, if and only if one of the following conditions holds:

c1) $\Omega_j^i=\emptyset$;

c2) $\Omega_j^i\ne \emptyset$, $i\ne j$, and for each $v_l^+ \in {\cal N}({\cal B}(H_\lambda), v_j^-)\backslash \{v_i^+\}$, every maximum matching of ${\cal B}(H_\lambda^{j_c})-\{v^+_l\}$ covers $\bigcup \nolimits_{k\in \Omega_j^i} {\cal G}_k^{j_c}$.\footnote{If ${\cal N}({\cal B}(H_\lambda), v_j^-)\backslash \{v_i^+\}=\emptyset$, the third item is automatically satisfied by property 4) of Definition \ref{DM-def} (the same below).}
\end{proposition}
%\footnote{If $N({\cal B}(H_\lambda), v_j^-)\backslash \{v_i^+\}=\emptyset$, the third item is always satisfied.}
%If vertex $v_i^+$ does not satisfy condition ({\bf $\star$})   Among $v_l^+\in \{v_l^+: [\bar A,\bar b]_{lj}\ne 0\}$, there is at most one vertex, being exactly $v_i^+$ and $i\ne j$, that does not satisfy condition ({\bf$\star$}).

Proposition \ref{nonzero-condition} gives the necessary and sufficient condition for the nonexistence of nonzero uncontrollable modes. Combining Propositions \ref{one-edge-principle}, \ref{graph-zero-mode}, and \ref{nonzero-condition} yields a necessary and sufficient condition for general single-input systems to be PSSC.

\begin{theorem}\label{main-result} Given $\bar A\in \{0,*,\times\}^{n\times n}$, $\bar b\in \{0,*,\times\}^{n\times 1}$,  $(\bar A, \bar b)$ is PSSC, if and only if $(\bar A, \bar b)$ is structurally controllable, and for each $\pi\doteq (i,j)\in {\cal N}_{\times}$, the following two conditions hold for the system $(\bar A^\pi, \bar b^\pi)$, recalling $(\bar A^\pi,\bar b^\pi)$ is defined in Proposition \ref{one-edge-principle}:

1) condition b1), condition b2), or condition b3) holds;

2) condition c1) or condition c2) holds.
\end{theorem}

We present some examples to illustrate Theorem \ref{main-result}. % Let us apply Theorem \ref{main-result} to

\begin{example}\label{exap2} Consider the system $(\bar A, \bar b)$ in Example \ref{SSC-PSSC}. In this system, ${\cal N}_\times=\{(3,2),(1,5)\}$. For $\pi=(3,2)$, the associated ${\cal B}([\bar A^\pi, \bar b^\pi])$, ${\cal B}(H_\lambda)$, and ${\cal B}(H_\lambda^{j_c})$ ($j=2$), as well as the DM-decomposition of ${\cal B}(H_\lambda^{j_c})$, are given respectively in Figs. \ref{examp1-a}, \ref{examp1-hl}, and \ref{examp1-b}. From Fig. \ref{examp1-a}, condition b2) is fulfilled, as the bipartite graph ${\cal B}([\bar A^\pi, \bar b^\pi])-\{v_4^+,v_2^-\}$ has a maximum matching with size $2$. From Fig. \ref{examp1-b}, $\Omega_2=\{2\}$ and $\Omega_2^3=\emptyset$, implying condition c1) is fulfilled. Similarly, for $\pi=(1,5)$, the associated ${\cal B}([\bar A^\pi, \bar b^\pi])$ and ${\cal B}(H_\lambda^{j_c})$ ($j=5$), as well as its DM-decomposition, are given respectively in Figs. \ref{examp1-a} and \ref{example1-d}. Fig. \ref{examp1-a} shows ${\cal N}({\cal B}([\bar A^\pi, \bar b^\pi]),v_5^-)\backslash \{v_1^+\}=\emptyset$, meaning condition b2) is satisfied.  Fig. \ref{example1-d} indicates $\Omega_5=\{2\}$ and $\Omega_5^1=\{2\}$, meanwhile, ${\cal N}({\cal B}(H_\lambda),v_5^-)\backslash \{v_1^+\}=\emptyset$. This means condition c2) is satisfied. As a consequence, $(\bar A, \bar b)$ is PSSC, which is consistent with the analysis in Example \ref{SSC-PSSC}. Further, suppose we change $(\bar A, \bar b)$ to {\small
\begin{equation}\label{modify_ab} \bar A=\left[
  \begin{array}{cccc}
    0 & 0 & 0 & 0 \\
    * & 0 & 0 & 0 \\
    0 & * & 0 & 0 \\
    \times & * & 0 & \times \\
  \end{array}
\right], \bar b=\left[
                  \begin{array}{c}
                    * \\
                    0 \\
                    0 \\
                    0 \\
                  \end{array}
                \right].
\end{equation}}Then, ${\cal N}_\times=\{(4,1),(4,4)\}$. For each $\pi\in {\cal N}_\times$, ${\cal B}([\bar A^\pi, \bar b^\pi])$ and the corresponding ${\cal B}(H_\lambda)$ are of the same form as Figs. \ref{examp1-a} and \ref{examp1-hl}, respectively. For $\pi=(4,4)$, ${\cal B}([\bar A^\pi, \bar b^\pi])$ satisfies condition b2), with the corresponding ${\cal B}(H_\lambda^{j_c})$ $(j=4)$ and its DM-decomposition given in Fig. \ref{examp2-add1}. It turns out that, $\Omega_j=\Omega_j^i=\{3\}$, and meanwhile, $i=j$. Therefore, neither condition c1) nor c2) is fulfilled, meaning $(\bar A, \bar b)$ in (\ref{modify_ab}) is not PSSC. This can be validated by using Theorem \ref{algebraic-condition}. Alternatively, we can obtain the same conclusion by inspecting that, for $\pi=(4,1)$, the corresponding ${\cal B}(H_\lambda^{j_c})$ ($j=1$; see Fig. \ref{examp2-add2}) does not satisfy condition c1) or c2). See, associated with ${\cal B}(H_\lambda^{j_c})$, $\Omega_j^i=\{1\}$, and there is a maximum matching of ${\cal B}(H_\lambda^{j_c})-\{v_4^+\}$ that does not cover ${\cal G}_1^{j_c}$.\hfill $\square$
\end{example}

\begin{figure} \centering
%\subfigure[${\cal B}({[\bar A^{\pi}, \bar b^\pi]}) $] { \label{examp1-a} %\ (\pi\in{\cal N}_\times)
%\includegraphics[width=0.27\columnwidth]{Ab2.eps}
%}
%\subfigure[${\cal B}(H_\lambda)$] { \label{examp1-hl}
%\includegraphics[width=0.28\columnwidth]{AbH2.eps}
%}
%\subfigure[${\cal B}(H_\lambda^{j_c}) \ (j=2)$] { \label{examp1-b}
%\includegraphics[width=0.37\columnwidth]{DM2.eps}
%}
%\subfigure[${\cal B}(H_\lambda^{j_c}) \ (j=5)$] { \label{example1-d}
%\includegraphics[width=0.38\columnwidth]{DM22.eps}
%}
%\subfigure[${\cal B}(H_\lambda^{j_c}) \ (j=4)$] { \label{examp2-add1}
%\includegraphics[width=0.37\columnwidth]{examp2-add1.eps}
%}
%\subfigure[${\cal B}(H_\lambda^{j_c}) \ (j=1)$] { \label{examp2-add2}
%\includegraphics[width=0.38\columnwidth]{examp2-add2.eps}
%}

\subfigure[${\cal B}({[\bar A^{\pi}, \bar b^\pi]}) $] { \label{examp1-a} %\ (\pi\in{\cal N}_\times)
\includegraphics[width=0.1944\columnwidth]{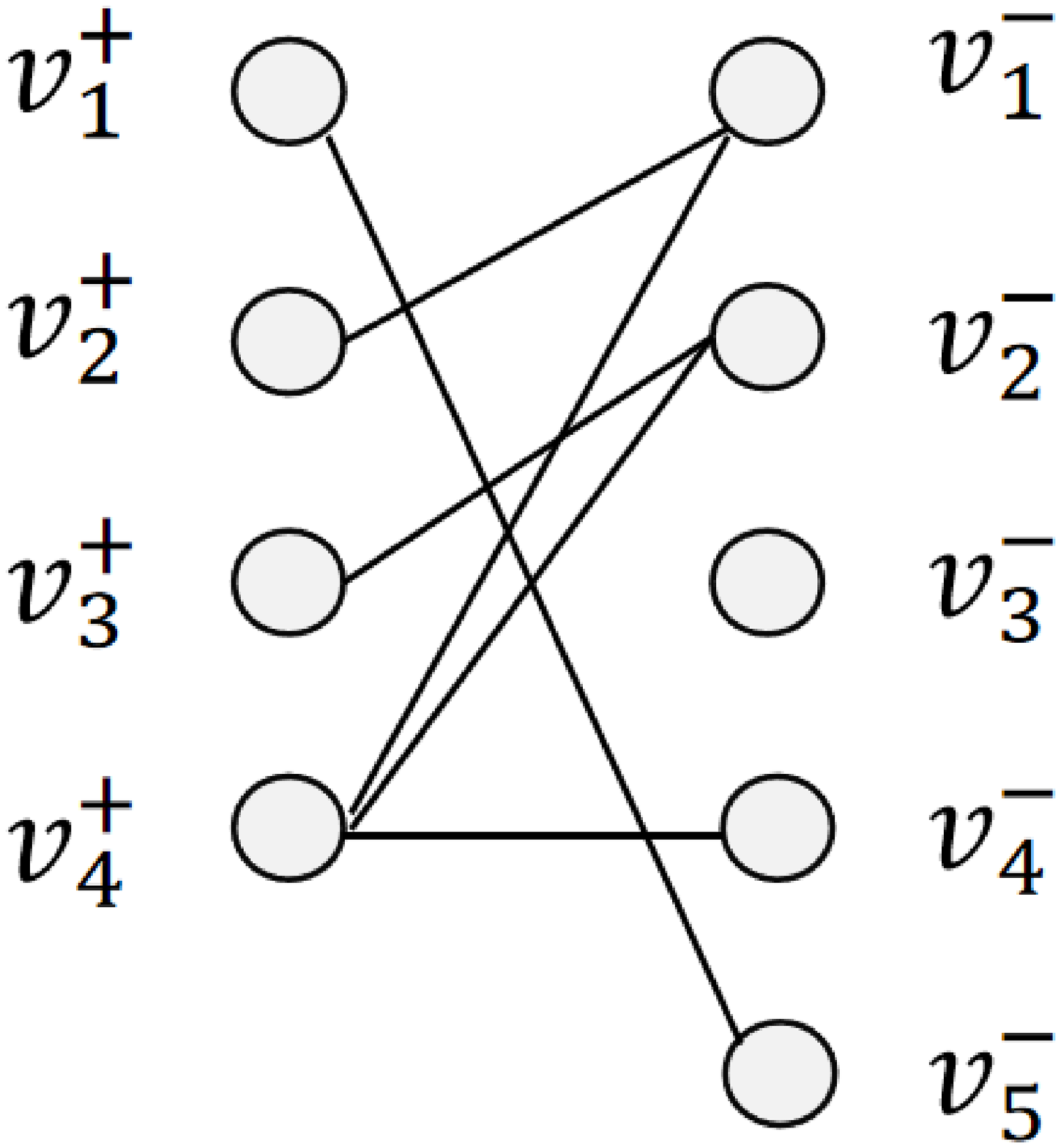}
}
\subfigure[${\cal B}(H_\lambda)$] { \label{examp1-hl}
\includegraphics[width=0.2016\columnwidth]{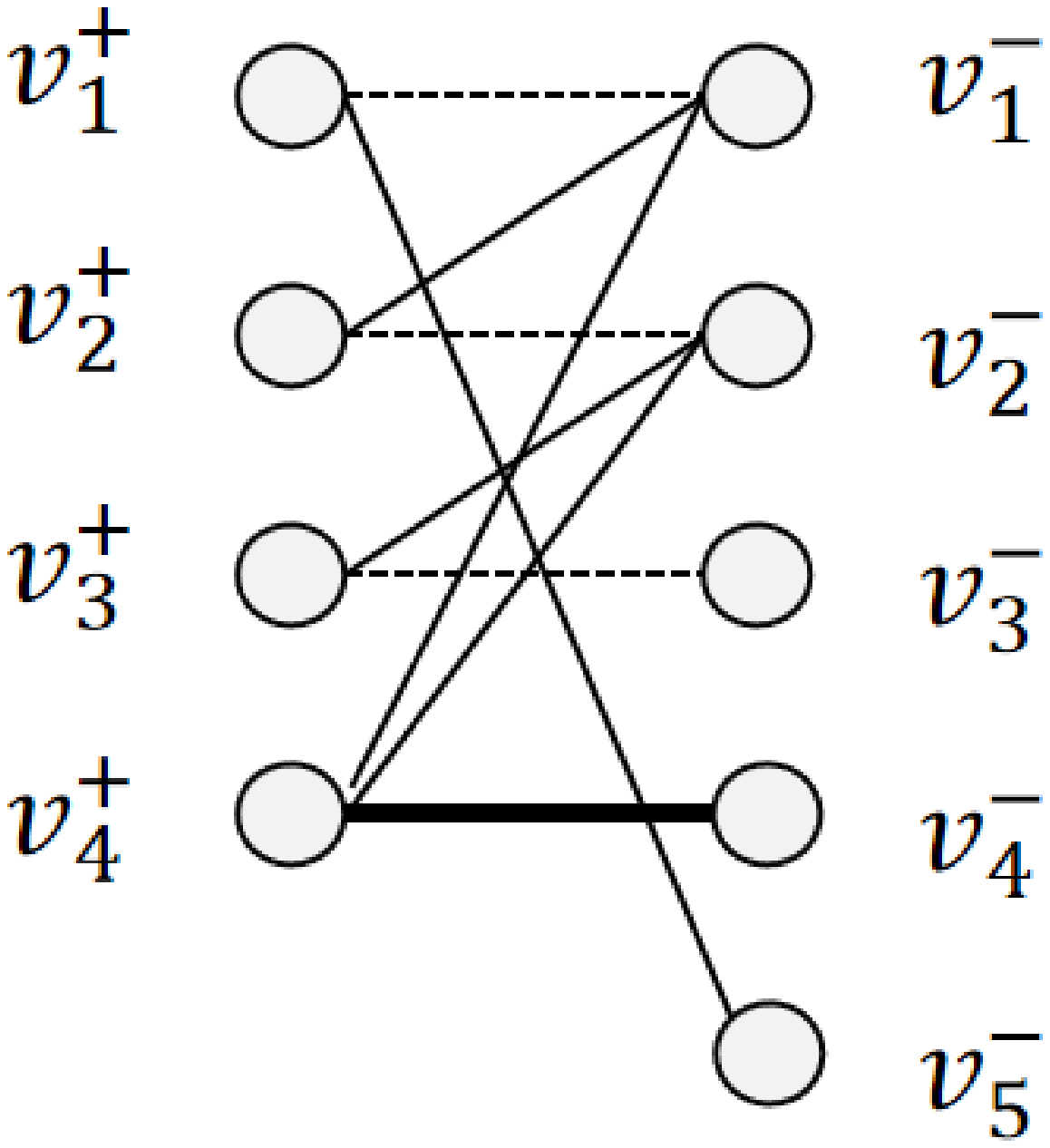}
}
\subfigure[${\cal B}(H_\lambda^{j_c}) \ (j=2)$] { \label{examp1-b}
\includegraphics[width=0.275\columnwidth]{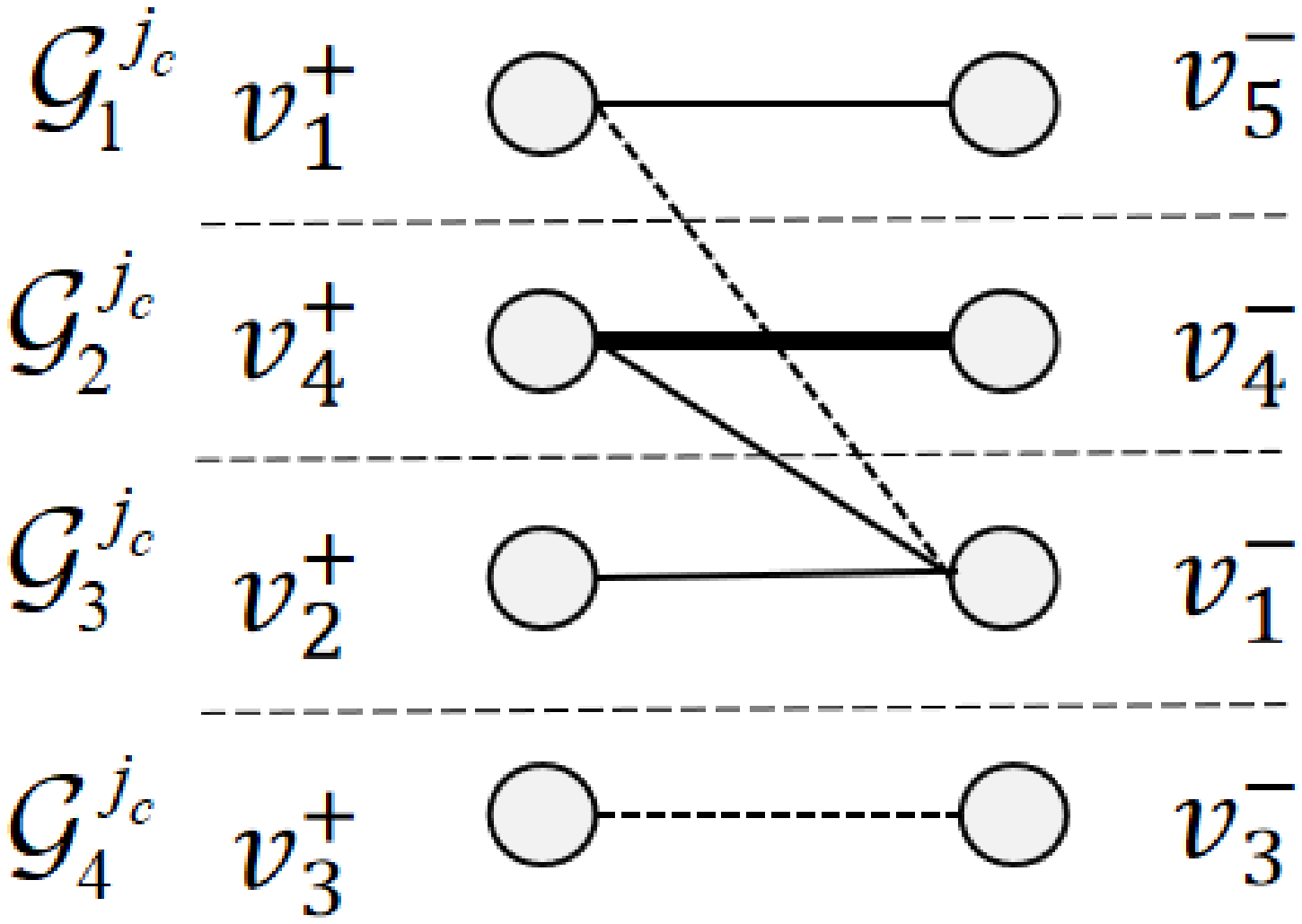}
}
\subfigure[${\cal B}(H_\lambda^{j_c}) \ (j=5)$] { \label{example1-d}
\includegraphics[width=0.2584\columnwidth]{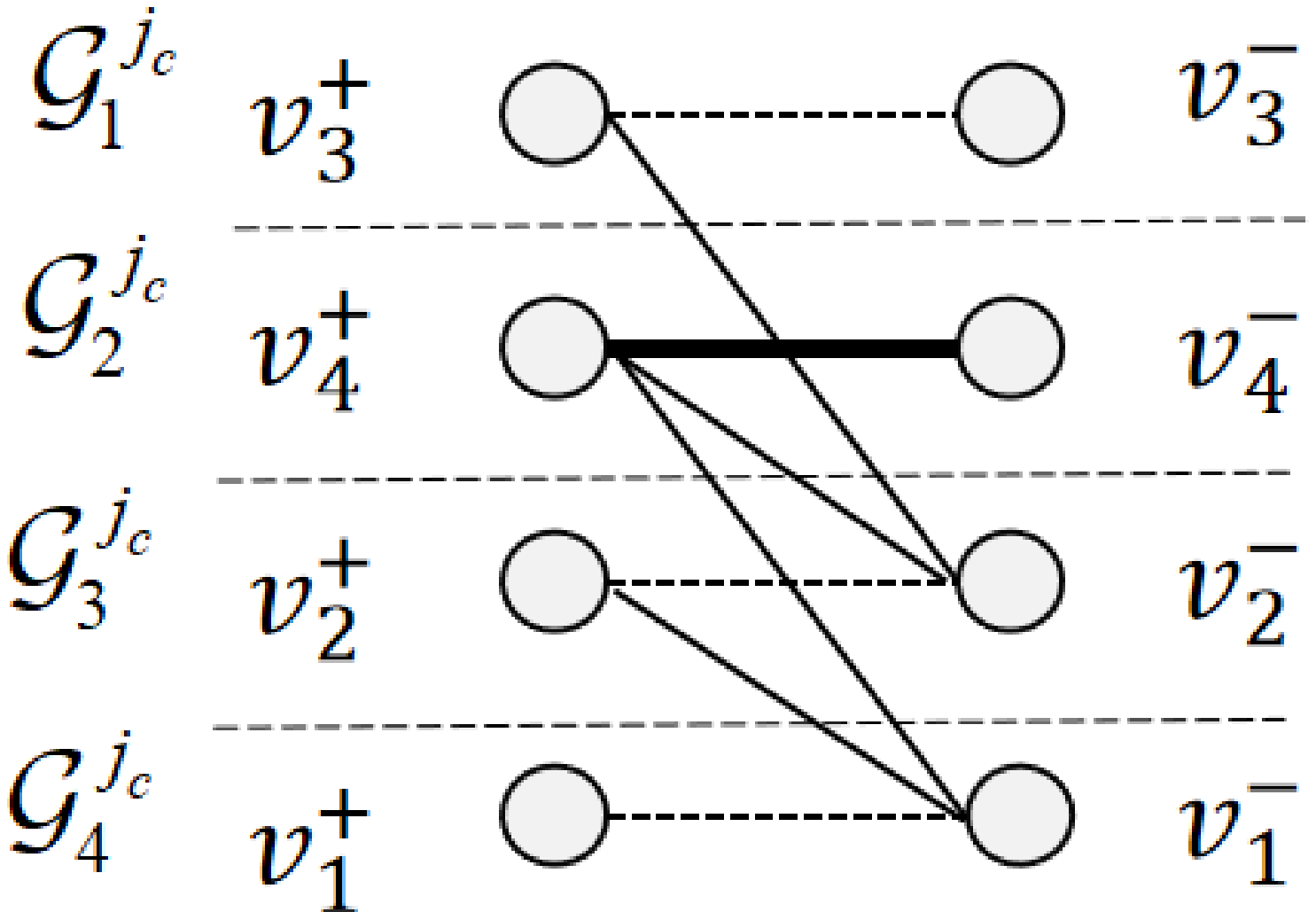}
}
\subfigure[${\cal B}(H_\lambda^{j_c}) \ (j=4)$] { \label{examp2-add1}
\includegraphics[width=0.2516\columnwidth]{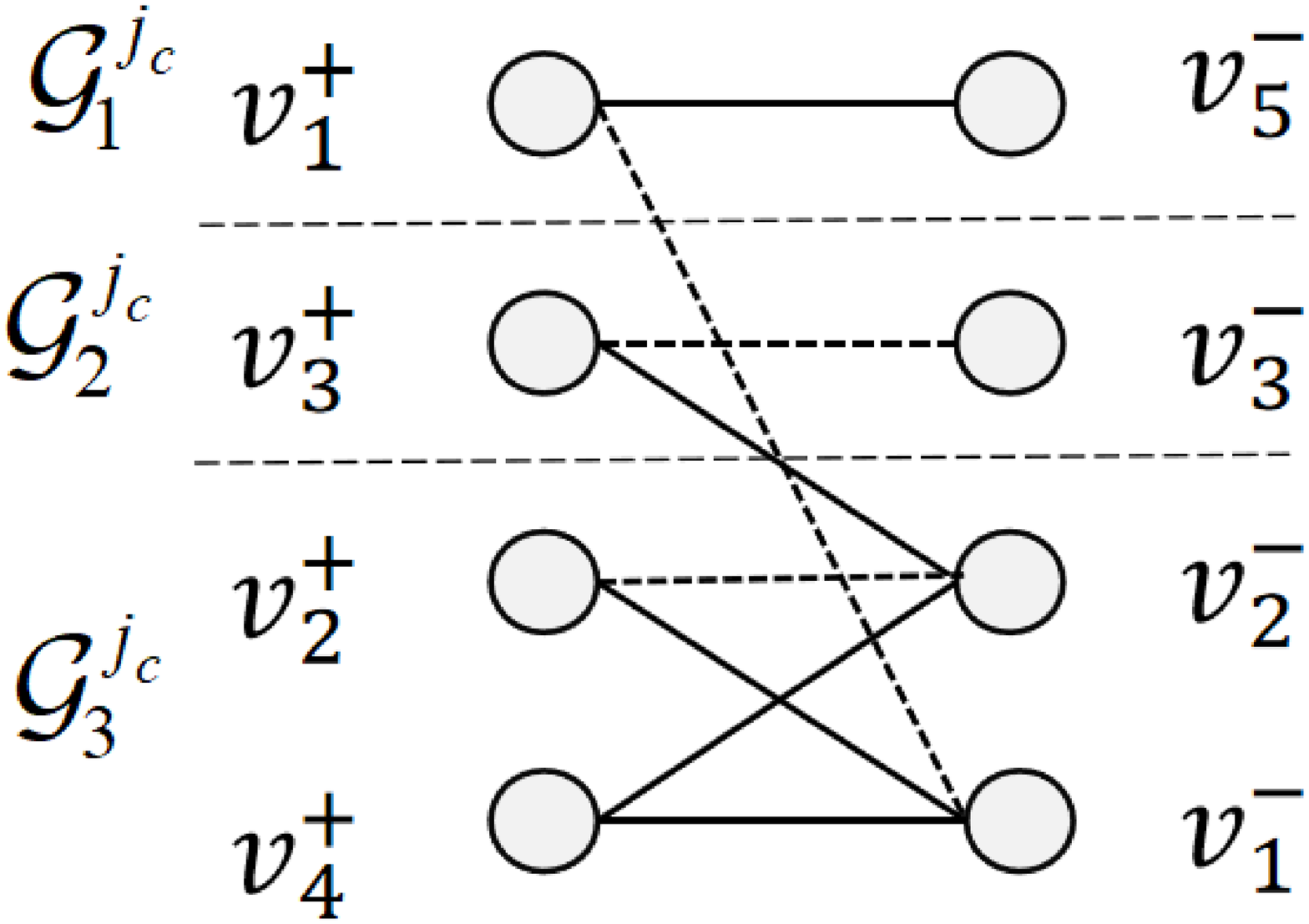}
}
\subfigure[${\cal B}(H_\lambda^{j_c}) \ (j=1)$] { \label{examp2-add2}
\includegraphics[width=0.2584\columnwidth]{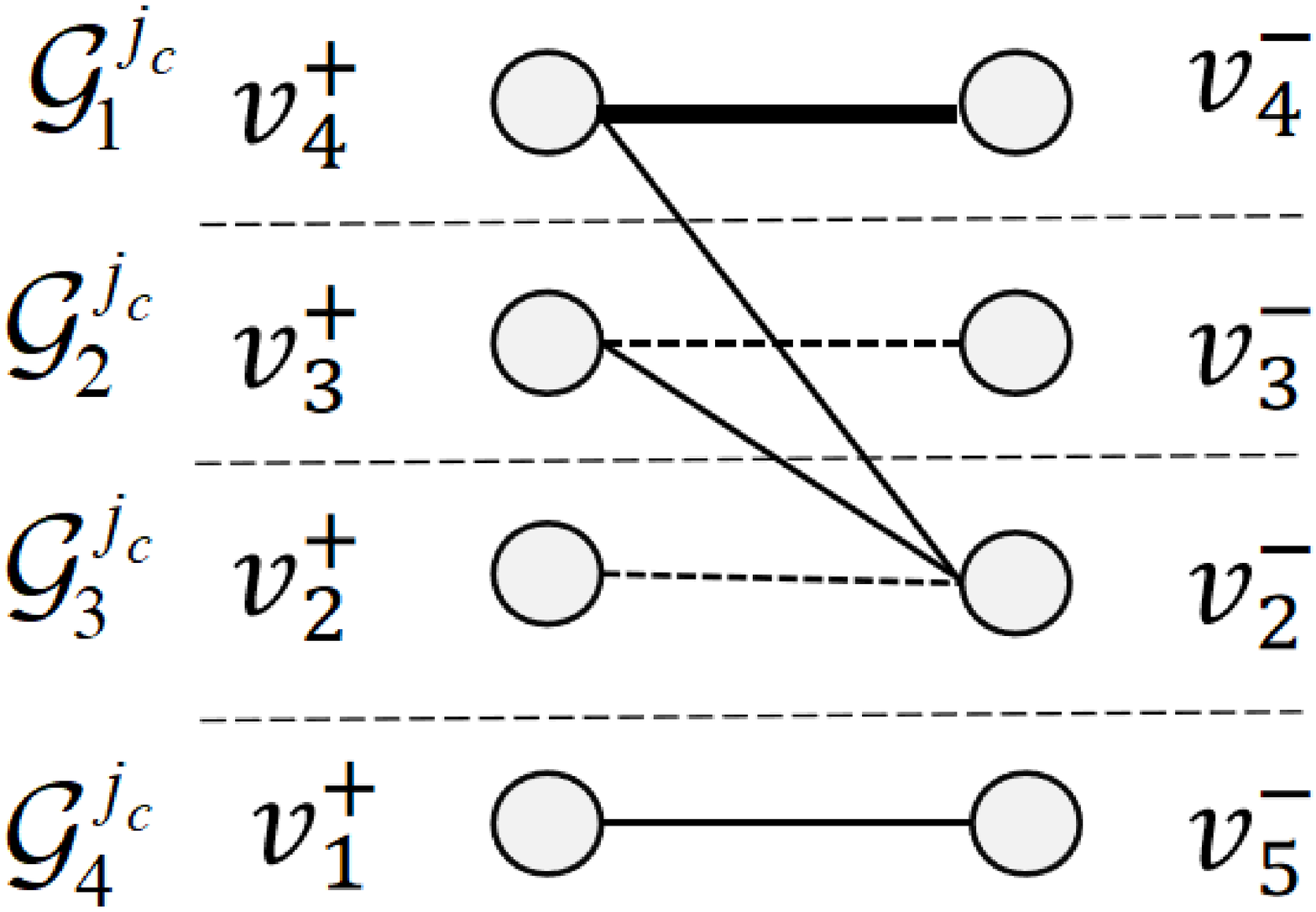}
}

\caption{Illustration of Theorem \ref{main-result} on systems (\ref{example-sys1}) and (\ref{modify_ab}). Bold edges represent self-loops, and dotted edges represent $\lambda$-edges that are not self-loops.}
\label{example-theo2-1}
\end{figure}

\begin{example}[Example \ref{examp-motivate} continuing] The system in Example \ref{examp-motivate} corresponds to the following structured matrices {\small
\begin{equation}\label{motivate_ab} \bar A=\left[
  \begin{array}{cccc}
    0 & \times & 0 & 0 \\
    * & 0 & 0 & 0 \\
    0 & 0 & 0 & \times \\
    * & 0 & 0 & 0 \\
  \end{array}
\right], \bar b=\left[
                  \begin{array}{c}
                    0 \\
                    * \\
                    0 \\
                    * \\
                  \end{array}
                \right].
\end{equation}}Following the similar process to Example \ref{exap2}, it turns out for $\pi=(1,2)$ and $\pi=(3,4)$, the corresponding systems $(\bar A^\pi, \bar b^\pi)$ both satisfy conditions b2) and c1). Hence, system (\ref{modify_ab}) is PSSC, consistent with the analysis in Example \ref{examp-motivate}. \hfill $\square$
\end{example}

%\begin{figure} \centering
%\subfigure[${\cal B}(H_\lambda^{j_c}) \ (j=4)$] { \label{examp2-add1}
%\includegraphics[width=0.40\columnwidth]{examp2-add1.eps}
%}
%\subfigure[${\cal B}(H_\lambda^{j_c}) \ (j=1)$] { \label{examp2-add2}
%\includegraphics[width=0.42\columnwidth]{examp2-add2.eps}
%}
%\caption{Illustration of Theorem \ref{main-result} on system (\ref{modify_ab}) or (\ref{}). The indications of different edge styles are the same as Fig. \ref{example-theo2-1}}
%\end{figure}
%the following three conditions hold (recalling $(\bar A^\pi,\bar b^\pi)$ is defined in Proposition \ref{one-edge-principle}):
%
%1) $(\bar A, \bar b)$ is structurally controllable;
%
%2) For each $\pi\doteq (i,j)\in {\cal N}_{\times}$, condition b1), condition b2), or condition b3) holds for the system $(\bar A^\pi, \bar b^\pi)$;
%
%3) For each $\pi\doteq (i,j)\in {\cal N}_{\times}$, condition c1) or condition c2) holds for the system $(\bar A^\pi, \bar b^\pi)$.

\subsection{Efficient verification of the proposed conditions} \label{sec-efficient-computation}
%In this subsection, we provide an efficient procedure to verify Theorem \ref{main-result}, described in Algorithm \ref{alg1}. Briefly, in this Algorithm, verifying conditions b1), b2) and b3) is implemented via the computation of bipartite maximum matching. Determining $\Omega_j$, $\Omega_j^i$, and whether each maximum matching of a bipartite graph covers a specific subgraph, is implemented via computing some minimum/maximum weight maximum matchings of weighted bipartite graphs.
%In this subsection, we show Theorem \ref{main-result} can be verified in polynomial time.

%In this subsection, we show Theorem \ref{main-result} can be verified efficiently mainly via (weighted) maximum matching computations, to be specific, in $O(|{\cal N}_{\times}|n^4)$ time.

Conditions b1), b2), and b3) can be verified directly via maximum matching computations on the associated bipartite graphs. As for conditions c1) and c2), first, $\gamma({\cal G}_k^{j_c})$ can be determined, as argued in \cite{full-version-tac}, via computing the minimum/maximum weight maximum matchings. Specifically, if we assign weight $1$ to each $\lambda$-edge of ${\cal G}_k^{j_c}$ and weight $0$ to the other edges, then by definition, $\gamma_{\max}({\cal G}_k^{j_c})$ equals the maximum weight of all maximum matchings of ${\cal G}_k^{j_c}$, and $\gamma_{\min}({\cal G}_k^{j_c})$ equals the minimum weight over all maximum matchings of ${\cal G}_k^{j_c}$. After determining $\Omega_j$, for a given $k\in \Omega_j$, assign weight to the edge $e$ of ${\cal B}(H_\lambda^{j_c})-\{v_i^+\}$ as follows
\begin{equation}\label{weight-procedure-1} w(e)=
\begin{cases} 1 & {\text{if}} \ e\in E_k\backslash \{(v_i^+,v^-_l): (v_i^+,v^-_l)\in E_{H_\lambda}\}  \\
0 &  \text{otherwise}.
\end{cases}\end{equation}
It is not hard to see, $k\in \Omega_j^i$, if and only if the minimum weight over all maximum matchings of the weighted ${\cal B}(H_\lambda^{j_c})-\{v_i^+\}$ is less than $|V_k^+|$. Indeed, if such a minimum weight is less than $|V_k^+|$, then there must be a maximum matching of ${\cal B}(H_\lambda^{j_c})-\{v_i^+\}$ that does not cover ${\cal G}_k^{j_c}$. On the other hand, if such a minimum weight is equal to $|V_k^+|$, then each maximum matching of ${\cal B}(H_\lambda^{j_c})-\{v_i^+\}$ should cover ${\cal G}_k^{j_c}$.
In such a manner, we can determine the set $\Omega_j^i$. Afterwards, if $i\ne j$ and $\Omega_j^i\ne 0$, for a given $v_l^+\in {\cal N}({\cal B}(H_\lambda),v_j^-)\backslash \{v_i^+\}$, to determine whether the third item of condition c2) is fulfilled,  we can adopt a similar manner to the preceding scenario, that is, assigning weight to the edge $e$ of ${\cal B}(H_\lambda^{j_c})-\{v^+_l\}$ as follows
\begin{equation}\label{weight-procedure-2} w(e)\!=\!
\begin{cases} 1 & {\text{if}}\  e\in \left(\bigcup \nolimits_{k\in \Omega_j^i} E_k\right)\backslash \{(v_l^+,v^-_{l'}): (v_l^+,v^-_{l'})\in E_{H_\lambda}\}  \\
0 &  \text{otherwise}.
\end{cases}\end{equation}
Then, similarly, it follows that, every maximum matching of ${\cal B}(H_\lambda^{j_c})-\{v^+_l\}$ covers $\bigcup \nolimits_{k\in \Omega_j^i} {\cal G}_k^{j_c}$, if and only if the minimum weight over all maximum matchings of the weighted ${\cal B}(H_\lambda^{j_c})-\{v^+_l\}$ equals $|\bigcup \nolimits_{k\in \Omega_j^i} V^+_{k}|$.  It is remarkable that determining whether $\Omega_j^i=\emptyset$ can be implemented in a similar manner, i.e., by replacing $\Omega_j^i$ in (\ref{weight-procedure-2}) with $\Omega_j$. %Indeed, if the minimum weight over all maximum matchings of the weighted ${\cal B}(H_\lambda^{j_c})-\{v^+_l\}$ is less than $|\bigcup \nolimits_{k\in \Omega_j^i} V^+_{k}|$, then there must a maximum matching of ${\cal B}(H_\lambda^{j_c})-\{v^+_l\}$ that does not cover $\bigcup \nolimits_{k\in \Omega_j^i} G_k^{j_c}$.

Let us figure out the computational complexity of the above procedure. Note that determining the maximum matching of a bipartite graph with $|V|$ vertices and $|E|$ edges incurs $O(\sqrt{|V|}|E|)$ time via the Hopcroft-Karp algorithm, and there are algorithms computing the maximum weighted matching in $O(|V|^3)$ \cite{DB_West_graph}. In addition, DM-decomposition has the same complexity as finding a maximum matching \cite{Murota_Book}. Verifying whether $(\bar A, \bar b)$ is structurally controllable can invoke the strongly-connected component decomposition and maximum matching algorithms, which incurs $O(n^{2.5})$.  Moreover, as analyzed above, for each $(\bar A^\pi,\bar b^\pi)$, conditions b1), b2) and b3) can be verified in $O(n\cdot n^{0.5}\cdot|E_X|)\to O(n^{3.5})$ time, and conditions c1) and c2) can be checked in time complexity at most{\small
$$\underbrace{O(n^{2.5})}_{{\rm DM-decomposition}}+\underbrace{O(n^3)}_{{\rm finding}\ \Omega_j}+\underbrace{O(n^4)}_{{\rm determining} \ \Omega_j^i}+ \underbrace{O(n^4)}_{\rm checking \ c2)}$$}$\to$ $O(n^4)$. To sum up, since there are $|{\cal N}_\times|$ $\times$ entries, the total time complexity of Theorem \ref{main-result} is at most $O(|{\cal N}_{\times}|n^4)$.

%one of the following conditions holds:
%
%1) $\Omega_j=\emptyset$;
%
%2) for each $k\in \Omega_j$, every maximum matching of ${\cal B}(H_\lambda^{j_c})-\{v^+_i\}$ contains $|V_k^+|$ edges of ${G_k^{j_c}}$;
%
%3)
%\end{proposition}

%, which incurs $O(n*n^{0.5}*||[\bar A,\bar b]||_0)\to O(n^{3.5})$ %and $[A-\lambda I, B][J_n,J_{n+m}\backslash \{j\}]$ ($j\in $)

\section{A special case for multi-input systems} \label{multi-case}
In this section, we consider PSSC for a special case in multi-input systems, that is, when there is only one $\times$ entry in $[\bar A, \bar B]$. Because of the property revealed in Proposition \ref{one-edge-principle}, such a case is enough to obtain necessary and sufficient conditions of PSSC for general single-input systems. {However, a similar property does not hold for multi-input systems,} for which the general case might need further inspection beyond the scope of this paper. %. The general case for multi-input systems might need further inspection beyond the scope of this paper. %, since ${\cal C}(A, B)$ is not square, which also makes the considered multi-input case more complicated and not a simple extension of the single-input one

% since for multi-input systems, ${\cal C}(A, B)$ is not square, the similar property does not hold. This also makes results in this section not a simple extension of the previous ones.}  The general case for multi-input systems might need further inspection beyond the scope of this paper. %{Since in the multi-input system, ${\cal C}(A, B)$ is not square,  results in this section are not simple extension of the previous section.} %We hope results in this subsection could  be a starting point towards more complicated scenarios. a similar property does not hold for

%Nevertheless,  detection and isolation of faults imposed on a single-edge of network systems has long been the research focus in the control community \cite{Commault2002ObserverbasedFD,Rahimian2015DetectionAI,zhang2020generic}, which is the starting point towards more complicated fault scenarios. We hope results in this subsection could function as a similar role.

In the rest of this section, recall $[\bar A, \bar B]$ is divided into $[\bar A, \bar B]=[\bar A_*,\bar B_*]+[\bar A_\times, \bar B_\times]$ in the way described in Definition \ref{PSSC-def}. First, the result below indicates the similar generic property in Proposition \ref{pro-generic} still holds for multi-input systems with a single $\times$ entry. %Let $[\bar A_*,\bar B_*]$ and $[\bar A_\times, \bar B_\times]$ be defined in the way described in Definition 3.

\begin{proposition}\label{pro-generic-multi} For a multi-input system $(\bar A, \bar B)$, assume that there is only one $\times$ entry in $[\bar A, \bar B]$. Then, either for almost all $[A_*,B_*]\in {\bf S}_{[\bar A_*,\bar B_*]}$, $(A_*+A_\times,B_*+B_\times)$ is controllable for each $[A_\times,B_\times]\in {\bf S}_{[\bar A_\times, \bar B_{\times}]}$, or for almost all $[A_*,B_*]\in {\bf S}_{[\bar A_*, \bar B_*]}$, there is a $[A_\times,B_\times]\in {\bf S}_{[\bar A_\times, \bar B_{\times}]}$ such that $(A_*+A_\times, B_*+B_\times)$ is uncontrollable.
\end{proposition}

The generic property presented above is characterized by PSSC of $(\bar A, \bar B)$. Next, similar to Proposition \ref{graph-zero-mode}, the following proposition gives the necessary and sufficient condition for the nonexistence of zero uncontrollable modes.

\begin{proposition}\label{zero-mode-multi}
Suppose $(\bar A, \bar B)$ is structurally controllable and $[\bar A, \bar B]$ contains only one $\times$ in its $(i,j)$th position.  For almost all $[A_*,B_*]\in {\bf S}_{[\bar A_*,\bar B_*]}$, there exist no $[A_\times, B_\times]\in {\bf S}_{[\bar A_\times, \bar B_\times]}$ and nonzero vector $q\in {\mathbb C}^n$ that satisfy $q^\intercal [A_*+A_\times, B_*+B_\times]=0$, if and only if one of the following conditions holds

%d1) ${\rm grank}([\bar A, \bar B][J_n,J_{n+m}\backslash \{j\}])=n$;
%
%d2) For each $k\in {\cal N}_*^j$, ${\rm grank}([\bar A, \bar B][J_n\backslash \{k\},J_{n+m}\backslash \{j\}])=n-2$, where ${\cal N}_{*}^j=\{k\in J_n\backslash \{i\}: [\bar A, \bar B]_{kj}= *\}$; %${\rm grank}([\bar A_*, \bar b_*][J_n\backslash \{i\},J_{n+1}\backslash \{j\}])=n-1$ and %${\rm grank}([\bar A_*, \bar b_*][J_n\backslash \{i\},J_{n+1}\backslash \{j\}])=n-1$, and for
%
%d3) ${\rm grank}([\bar A, \bar B][J_n\backslash \{i\},J_{n+m}\backslash \{j\}])=n-2$.

d1) ${\cal B}([\bar A, \bar B])$ contains a maximum matching that does not match $v_j^-$;

d2) For each $v_k^+\in {\cal N}({\cal B}([\bar A, \bar B]),v_j^-)\backslash \{v_i^+\}$ (if exists),\\ ${\rm mt}({\cal B}([\bar A, \bar B])-\{v_k^+,v_j^-\})=n-2$;

d3) ${\rm mt}({\cal B}([\bar A, \bar B])-\{v_i^+,v_j^-\})=n-2$.
\end{proposition}
%Then, the property stated in Proposition \ref{pro-generic} still holds for $[\bar A_*,\bar B_*]$ and $[\bar A_\times, \bar B_\times]$.
%Suppose the unique $\times$ entry in $[\bar A, \bar B]$ is in its $(i,j)$th position.

Suppose $(\bar A, \bar B)$ is structurally controllable and $[\bar A, \bar B]$ contains only one $\times$ in its $(i,j)$th position. Let
$[\tilde A, \tilde B]$ be a generic realization of $[\bar A, \bar B]$. Define a generic matrix pencil $H_{\lambda}=[\tilde A-\lambda I, \tilde B]$, and let $H_{\lambda}^{j_c}=H_{\lambda}[J_n,J_{n+1}\backslash\{j\}]$. Let ${\cal B}(H_\lambda)=(V^+,V^-,E_{H_\lambda})$ and ${\cal B}(H^{j_c}_\lambda)$ be the bipartite graphs associated with $H_{\lambda}$ and $H^{j_c}_\lambda$, respectively, defined in the same way as in Section \ref{single-case}.  {Note compared with the single-input case, the essential difference is that $|V^-\backslash \{v_j^-\}|=n+m-1\ge |V^+|$, which results in that there are horizontal or vertical tails in DM-decomposing ${\cal B}(H^{j_c}_\lambda)$.}  Owing to the structural controllability of $(\bar A, \bar B)$, a trivial extension of \citep[Lem 4]{full-version-tac} shows ${\rm mt}({\cal B}(H_\lambda^{j_c}))=n$. Consequently, by Definition \ref{DM-def}, there is only a horizontal tail in the DM-decomposition of ${\cal B}(H^{j_c}_\lambda)$ ($m>1$). Let ${\cal G}_k^{j_c}=(V_k^+,V_k^-,E_k)$ ($k=0,1,...,d$) be the DM-components of ${\cal B}(H^{j_c}_\lambda)$. {The following intermediate result is crucial for the subsequent derivations.}  %${\cal B}(H_\lambda)=(V^+,V^-,E)$,

\begin{lemma} \label{horizontal-no-root} If $m>1$, there is generically no nonzero $\lambda$ that can make $H^{j_c}_\lambda[V^+_0,V^-_0]$ row rank deficient.
\end{lemma}

%$0\notin \Omega_j$.

Moreover, associated with ${\cal B}(H_\lambda^{j_c})$ and ${\cal G}_k^{j_c}$ ($k=0,1,...,d$), let $\Omega_j$ and $\Omega_j^l$ ($v_l^+\in V^+$) be defined in the same way as (\ref{important-set}) and (\ref{important-subset}), respectively. Particularly, Lemma \ref{horizontal-no-root} implies { $H^{j_c}_\lambda[V^+_0,V^-_0]$ would contribute no nonzero $\lambda$ that can make $H^{j_c}_\lambda$ row rank deficient (thus $0\notin {\Omega_j}$)}. We have the following proposition, providing a necessary and sufficient condition for the nonexistence of nonzero uncontrollable modes. %, in a form analogous to Proposition \ref{nonzero-condition}.

\begin{proposition}\label{nonzero-mode-multi}
Suppose $(\bar A, \bar B)$ is structurally controllable and $[\bar A, \bar B]$ contains only one $\times$ entry in its $(i,j)$th position.   For almost all $[A_*,B_*]\in {\bf S}_{[\bar A_*,\bar B_*]}$, there exist no $[A_\times, B_\times]\in {\bf S}_{[\bar A_\times, \bar B_\times]}$, nonzero complex number $z$ and nonzero vector $q\in {\mathbb C}^n$ that satisfy $q^\intercal [A_*+A_\times-z I, B_*+B_\times]=0$, if and only if one of the following conditions holds

e1) $\Omega_j^i=\emptyset$;

e2) $\Omega_j^i\ne \emptyset$, $i\ne j$, and for each $v_l^+ \in {\cal N}({\cal B}(H_\lambda), v_j^-)\backslash \{v_i^+\}$, every maximum matching of ${\cal B}(H_\lambda^{j_c})-\{v^+_l\}$ covers $\bigcup \nolimits_{k\in \Omega_j^i} {\cal G}_k^{j_c}$.
\end{proposition}

Combining Propositions \ref{zero-mode-multi} and \ref{nonzero-mode-multi} yields a necessary and sufficient condition for PSSC of $(\bar A, \bar B)$ with a single $\times$ entry.

\begin{theorem}\label{theorem-multi-input} Suppose $(\bar A, \bar B)$ contains only one $\times$ entry in its $(i,j)$th position. $(\bar A, \bar B)$ is PSSC, if and only if: i) $(\bar A, \bar B)$ is structurally controllable, ii) Condition d1), d2) or d3) holds, and iii) Condition e1) or e2) holds.
\end{theorem}

Similar to the single-input case, Theorem \ref{theorem-multi-input} can be verified in polynomial time mainly via the (weighed) maximum matching computations. Specifically, following a similar manner to Section \ref{sec-efficient-computation}, it can be found the total complexity of Theorem \ref{theorem-multi-input} is at most $O(n(n+m)^3)$.

{
\begin{remark}
Although presented in an analogous form to the single-input case, results in this section are not simple extensions of the previous section. As shown in our derivations, since ${\cal C}(A,B)$ is not square, the proof for genericity needs to consider multiple $n\times n$ submatrices of ${\cal C}(A,B)$. Besides, as $H_\lambda^{j_c}$ is no longer square, we have to consider the horizontal tail of the DM-components of ${\cal B}(H_{\lambda}^{j_c})$.
\end{remark}}
%More precisely, following a similar manner to Section \ref{sec-efficient-computation}, it can be found verifying condition i) incurs $O(n(n+m)^{1.5})$, and verifying condition ii) costs $O(n\cdot (n+m)^{0.5}\cdot|E_X|)\to O(n^2(n+m)^{1.5})$, while verifying condition iii) takes $O(n(n+m)^3)$. Therefore, the total complexity of Theorem \ref{theorem-multi-input} is at most $O(n(n+m)^3)$.

While Theorem \ref{theorem-multi-input} is devoted to the single $\times$ entry case, it can provide some necessary conditions for PSSC of more general cases. Specifically, it is easy to see, for $(\bar A, \bar B)$ to be PSSC, by preserving arbitrary one of its $\times$ entries and changing the remaining $\times$ entries to $*$, the obtained structured system should be PSSC, i.e., satisfying the conditions in Theorem \ref{theorem-multi-input}; otherwise, $(\bar A, \bar B)$ cannot be PSSC by Proposition \ref{pro-generic-multi}.

%\begin{remark} While Theorem \ref{theorem-multi-input} is devoted to the single $\times$ entry case, it can provide some necessary conditions for PSSC of more general cases. Specifically, it is easy to see, for $(\bar A, \bar B)$ to be PSSC, by preserving arbitrary one of its $\times$ entries and changing the remaining $\times$ entries to $*$, the obtained structured system should be PSSC, i.e., satisfying the conditions in Theorem \ref{theorem-multi-input}.
%\end{remark}

\begin{example} Consider $(\bar A,\bar B)$ as{\small
\begin{equation}\label{example-multi}\bar A=\left[
     \begin{array}{cccccc}
       0 & 0 & 0 & 0 & 0 & 0 \\
       0 & * & * & 0 & 0 & 0 \\
       0 & 0 & 0 & * & * & 0 \\
       0 & 0 & 0 & 0 & \times & 0 \\
       0 & 0 & 0 & 0 & 0 & * \\
       0 & * & 0 & 0 & 0 & *\\
     \end{array}
   \right],  \bar B=\left[
                \begin{array}{cc}
                  * & 0 \\
                  * & 0 \\
                  0 & 0 \\
                  0 & 0 \\
                  0 & * \\
                  0 & * \\
                \end{array}
              \right],
\end{equation}}which is structurally controllable. The associated ${\cal B}([\bar A, \bar B])$ and ${\cal B}(H_\lambda^{j_c})$ (and its DM-decomposition, $j=5$) are given in Figs. \ref{examp3-a} and \ref{examp3-b}.
From them, we know ${\cal B}([\bar A, \bar B])-\{v^+_3,v^-_5\}$ has a maximum matching with size $4$, and $\Omega_j=\emptyset$. This means the conditions of Theorem \ref{theorem-multi-input} are satisfied. Hence, $(\bar A, \bar B)$ is PSSC. Further, it can be found that, by replacing arbitrary one of the indeterminate entries of $(\bar A, \bar B)$ with a $\times$ entry and changing the remaining ones to $*$, the obtained structured system is still PSSC. This is consistent with the fact that $(\bar A, \bar B)$ is actually SSC (c.f. \citep[Theo 4]{jarczyk2011strong}). \hfill $\square$
\end{example}

\begin{figure} \centering
\subfigure[${\cal B}({[\bar A, \bar B]})$] { \label{examp3-a}
\includegraphics[width=0.1925\columnwidth]{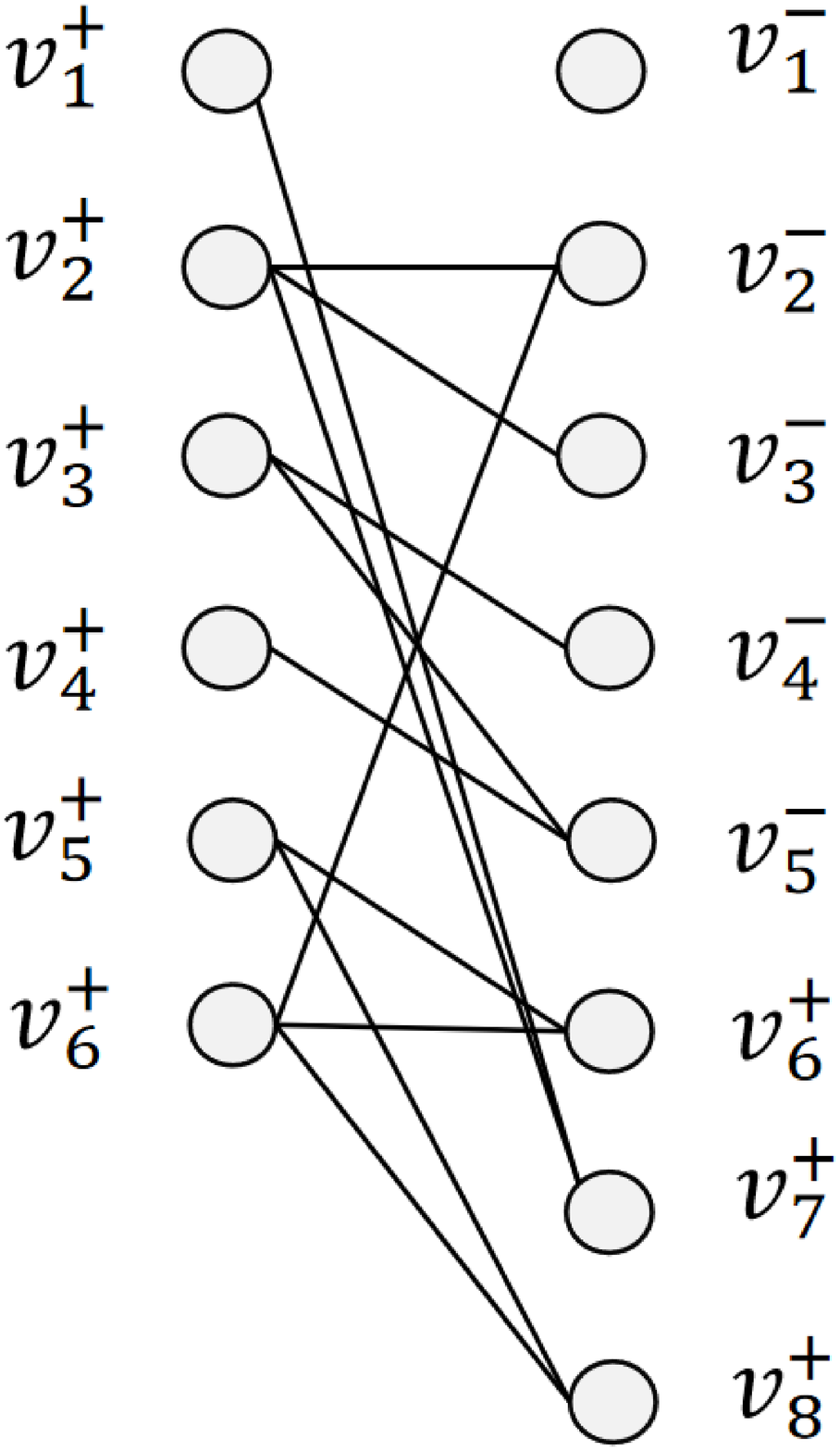}
}
\subfigure[${\cal B}(H_\lambda^{j_c}) \ (j=5)$] { \label{examp3-b}
\includegraphics[width=0.2834\columnwidth]{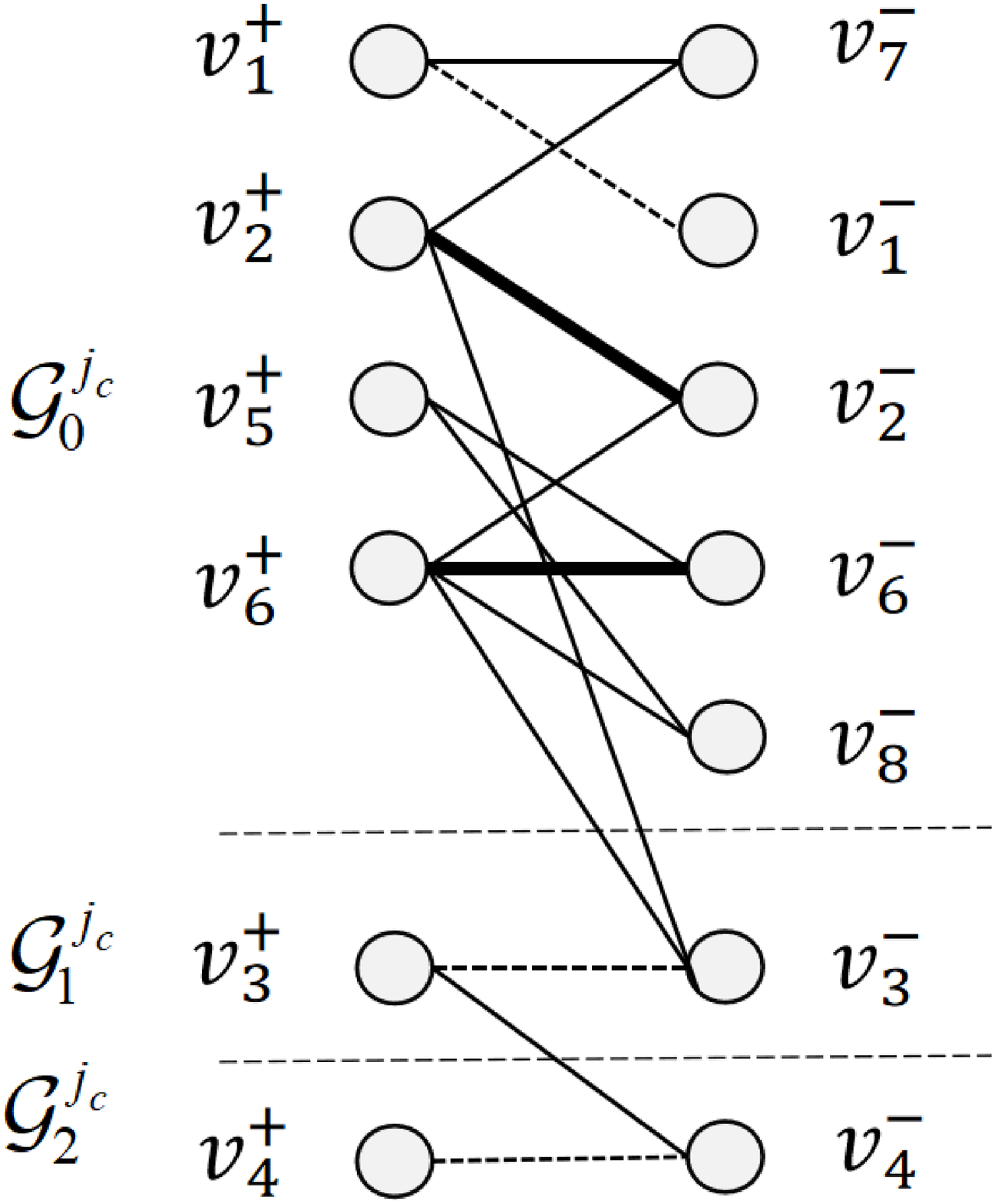}
}
\caption{Illustration of Theorem \ref{theorem-multi-input} on systems (\ref{example-multi}). The indications of different edge styles are the same as Fig. \ref{example-theo2-1}.}
\label{example-theo3-multi}
\end{figure}

\vspace{-1em}
\section{Implications {for} the existing SSC theory} \label{SSC-corollary-section}
In this section, we point out the proposed PSSC criteria in the special case can provide new graph-theoretic conditions for SSC, even restricted to the real field. Further, we demonstrate the new conditions can provide a classification of edges (or indeterminate entries) with respect to system controllability.

%contain the existing criteria for SSC as special cases.

As mentioned earlier, when there is no $*$ entry in $[\bar A, \bar b]$, Theorem \ref{main-result} collapses to the criterion for SSC (in the complex field). With this idea, the following corollary provides a new necessary and sufficient condition for SSC in terms of (weighted) maximum matchings over the system bipartite graph representation.

\begin{corollary}\label{SSC-corollary} Given $\bar A\in \{0,\times\}^{n\times n}$, $\bar b\in \{0,\times \}^{n\times 1}$, $(\bar A, \bar b)$ is SSC {\bf in the real field} (i.e., $(A,b)$ is controllable for all {\bf real-valued} $[A, b]\in {\bf S}_{[\bar A, \bar b]}$), if and only if $(\bar A, \bar b)$ is structurally controllable, and for each $\pi \in {\cal N}_{\times}$, $(\bar A^\pi,\bar b^\pi)$ satisfies: i) at least one of conditions b1), b2) or b3) holds; and ii) condition c1) or condition c2) holds.
\end{corollary}

{Although it is not hard to expect that, necessary and sufficient conditions for SSC in the real field and the complex field should have the same form (c.f. \cite{bowden2012strong}), we have provided a self-contained proof for the corollary above from the developed PSSC theory (see the Appendix)}.  Compared to \citep[Theo 5]{chapman2013strong}, Corollary \ref{SSC-corollary} is not appealing in terms of computational complexity. Nevertheless, since Corollary \ref{SSC-corollary} is an entry-wise criterion, it does provide some deep insight into the role of each edge of ${\cal G}(\bar A,\bar b)$ in system controllability. For description convenience, for $[\bar A, \bar b]\in \{0,\times\}^{n\times(n+1)}$, given $\pi\in {\cal N}_\times$, let $\bar p_\pi$ be the parameter for the $\pi$th $\times$ entry of $[\bar A, \bar b]$, and $\bar p_{\pi^{c}}$ the vector consisting of parameters for the remaining $\times$ entries. Given that $(\bar A, \bar b)$ is structurally controllable, for any $\pi=(i,j)\in {\cal N}_\times$, depending on what conditions in Theorem \ref{main-result} are satisfied for $(\bar A^\pi,\bar b^\pi)$, the edge $(x_j,x_i)$ can be classified into: % Particularly
%\begin{center}
\begin{itemize}
\item    Critical edge: there is a nonzero (complex) value for $\bar p_\pi$ making the corresponding realization uncontrollable, for almost all $\bar p_{\pi^{c}}\in \bar {\mathbb C}^{n_\times -1}$, if $(\bar A^\pi,\bar b^\pi)$ does not satisfy condition 1) or condition 2) of Theorem \ref{main-result};

\item   Stable edge:  there is no nonzero value for $\bar p_\pi$ that can make the corresponding realization uncontrollable, for almost all $\bar p_{\pi^{c}}\in \bar{\mathbb C}^{n_\times -1}$, if $(\bar A^\pi,\bar b^\pi)$ satisfies conditions 1) and 2) of Theorem \ref{main-result};

%\item    Robust edge: for almost all $\bar p_{\pi^{c}}\in \bar{\mathbb C}^{n_\times -1}$, the corresponding realization is controllable for all $\bar p_\pi\in {\mathbb C}$, if condition b1) or b3) is fulfilled for $(\bar A^\pi,\bar b^\pi)$, and meanwhile condition c1) is satisfied.
\end{itemize}
The above classification rule is immediate from the definitions of PSSC and Theorem \ref{main-result}.

\begin{example} Consider $(\bar A, \bar b)$
%as
%\begin{equation} \label{sys-edge-classification}
%\bar A=\left[
%  \begin{array}{cccc}
%    0 & 0 & \times & \times \\
%    \times & 0 & 0 & 0 \\
%    0 & \times & 0 & 0 \\
%    \times & \times & 0 & \times \\
%  \end{array}
%\right], \bar b=\left[
%                  \begin{array}{c}
%                    \times \\
%                    0 \\
%                    0 \\
%                    0 \\
%                  \end{array}
%                \right],
%\end{equation}
that is obtained from system (\ref{SSC-PSSC}) by replacing all its indeterminate entries with $\times$ entries. Corollary \ref{SSC-corollary} yields that $(\bar A, \bar b)$ is not SSC. Further, a byproduct of Corollary \ref{SSC-corollary} on this system is the following classifications for its edges (which can be obtained from Example \ref{exap2}): (a) critical edges: $(x_1,x_2)$, $(x_1,x_4)$, $(x_2,x_4)$ and $(x_4,x_4)$; (b) stable edges:  $(x_2,x_3)$, and $(x_5,x_1)$.
%\begin{itemize}
%\item    Critical edges: $(x_1,x_2)$, $(x_1,x_4)$, $(x_2,x_4)$ and $(x_4,x_4)$;
%
%\item   Stable edges:  $(x_2,x_3)$, and $(x_5,x_1)$;
%
%%\item    Robust edges: $(x_3,x_1)$ and $(x_4,x_1)$.
%\end{itemize}
It is easy to validate the above assertion by noting that the determinant of a generic realization of $(\bar A, \bar b)$ is exactly the right-hand side of (\ref{det-expression}). \hfill $\square$
\end{example}

\section{Conclusions}
In this paper, a new controllability notion, named PSSC, has been proposed for linear systems, with the aim to extend the existing SSC and bridge the gap between structural controllability and SSC. %This new notion is defined on a structured system as the property that, for {\emph{almost all}} values of a prescribed subset of its indeterminate entries, the corresponding system realization is controllable for {\emph{all}} nonzero values of the remaining indeterminate entries.
 Algebraic and bipartite graph-theoretic necessary and sufficient conditions are given for single-input systems to be PSSC, the latter of which can be verified efficiently. Extension to the multi-input case on a special case is also given. Further, it is demonstrated the established results for PSSC in the single-input case could provide a new graph-theoretic criterion for the conventional SSC. In the future, we plan to extend the previous results to the general multi-input case, and to investigate PSSC in the real field systematically. It is also interesting to enumerate the number of PSSC networks statistically.
% the indeterminate entries of a structured system are divided into two categories, generic entries more in depth. %exquisitely. % It is shown PSSC generalizes the generic property embedded in the conventional structural controllability.

\section*{Appendix: Proofs of the technical results}
{\bf{Proof of Theorem \ref{algebraic-condition}:}} Sufficiency: Let $p\doteq (p_1,...,p_{n_*})$. Since $f(p)$ is a nonzero polynomial, $p\in {\mathbb C}^{n_*}$ making  $f(p)=0$ has zero measure in ${\mathbb C}^{n_*}$. Therefore, in case $\bar p_i\ne 0$ ($i=1,...,n_\times$) and $f(p)\ne 0$, it follows $f(p)\prod \nolimits_{i=1}^{n_\times} \bar p_i^{r_i}\ne 0$, indicating the corresponding realization is always controllable.

Necessity: The necessity of $\det {\cal C}(\tilde A, \tilde b)\ne 0$ is obvious.  If $\det {\cal C}(\tilde A, \tilde b)\ne 0$ but the remaining condition is not satisfied, then there is a $\bar p_i$ ($1\le i \le n_{\times}$) that exists in two different monomials with different degrees (including zero) for $\bar p_i$ (the degree of $\bar p_i$ is the exponent of $\bar p_i$) in $\det {\cal C}(\tilde A, \tilde b)$. Let $\bar p'=(\bar p_1,...,\bar p_{i-1},\bar p_{i+1},...,\bar p_{n_\times})$ and $p'=(p_1,...,p_{n_*}) \cup \bar p'$.  In this case, write $\det {\cal C}(\tilde A, \tilde b)$ as a polynomial of $\bar p_i$ as
\begin{equation}\label{non-PSSC-form} \det {\cal C}(\tilde A, \tilde b)=f_r\bar p_i^r+\cdots+f_1\bar p_i+ f_0\doteq g(p';\bar p_i),\end{equation}
in which the coefficients $f_j$ ($j=0,...,r$) are polynomials of $p'$, and $f_r$ as well as another $f_j$ ($0\le j < r$) is not identically zero. Consider the set
${\cal P}_1=\{(p_1,...,p_{n_*})\in {\mathbb C}^{n_*}: \exists \bar p' \in {\mathbb C}^{n_\times-1}, {\rm{s.t.}}\  f_r\ne 0, f_j\ne 0, \prod \nolimits_{k=1,k\ne i}^{n_\times}\bar p_{k}\ne 0\}$. Obviously, the complement of ${\cal P}_1$ in ${\mathbb C}^{n_*}$ has zero measure, as $\{p'\in {\mathbb C}^{n_*+n_\times-1}: f_r\ne 0, f_j\ne 0, \prod \nolimits_{k=1,k\ne i}^{n_\times}\bar p_{k}\ne 0\}$ has full dimension in ${\mathbb C}^{n_*+n_\times-1}$. Note in case $f_r\ne 0$ and $f_j\ne 0$, $g(p';\bar p_i)$ has at least one nonzero root for $\bar p_i$ (as otherwise $g(p';\bar p_i)=f_r\bar p_i^r$).  Therefore, for all $(p_1,...,p_{n_*})\in {\cal P}_1$, there exists $(\bar p_1,...,\bar p_{n_\times})\in \bar {\mathbb C}^{n_\times}$ such that  $g(p';\bar p_i)=0$, making the obtained realization uncontrollable.

{\bf{Proof of Proposition \ref{pro-generic}:}} The statement follows directly from the proof of Theorem \ref{algebraic-condition}. To be specific, for a given $(\bar A, \bar b)$, the first case emerges if $(\bar A, \bar b)$ is PSSC, while the second case emerges if $(\bar A, \bar b)$ is not PSSC.

%if $\Gamma(c,s)$ ($\ne 0$) is of the form $\Gamma(c,s)=P(c)s^r$, where $r\ge 0$ and $P(c)$ is a polynomial of $c$, then for all $\{c\in {\mathbb C}^{n_*}: P(c)\ne 0\}$ and all $s\ne 0$, all . %The case with $(\bar A, \bar B)$ being structurally uncontrollable  trivially
% recalling $n^*$ is the number of $*$ entries,

{\bf{Proof of Proposition \ref{one-edge-principle}:}} %We will prove that,  $(\bar A, \bar b)$ is not PSSC, if and only if there is a pair $(i,j)\in {\cal N}_\times$, such that $(\bar A^\pi,\bar b^\pi)$ is not PSSC. For the one direction,
By Theorem \ref{algebraic-condition}, if $(\bar A, \bar b)$ is PSSC, then for its generic realization $(\tilde A, \tilde b)$, $\det{\cal C}(\tilde A, \tilde b)$ has the form of (\ref{formal-form}). Let $[\tilde A^\pi,\tilde b^\pi]$ be the generic realization of $[\bar A^\pi,\bar b^\pi]$.  It is easy to see, for every $\pi\in {\cal N}_\times$,  $\det {\cal C}(\tilde A^\pi,\tilde b^\pi)$ then has the form of (\ref{formal-form}), which indicates $(\bar A^\pi,\bar b^\pi)$ is PSSC.

On the other hand, suppose $(\bar A^\pi,\bar b^\pi)$ is PSSC, $\forall \pi\doteq (i,j)\in {\cal N}_\times$. Let $\bar p_{\pi}$ be the parameter for the $(i,j)$th entry of $[\bar A, \bar b]$. By Theorem \ref{algebraic-condition}, for each $\pi\in {\cal N}_{\times}$, $\det {\cal C}(\tilde A^\pi,\tilde b^\pi)$ has a factor $\bar p_{\pi}^{r_{\pi}}$ for some $r_{\pi}\ge 0$, and any other factor containing $\bar p_{\pi}$ does not exist.  Consequently, $\det {\cal C}(\bar A, \bar b)$ must have a form of
(\ref{formal-form}). This means, $(\bar A, \bar b)$ is PSSC.

%{\bf Proof of Proposition \ref{zero-condition}:} %To prove this proposition, we need the following intermediate result.
%
%
%({\bf{Proof counting}}) We are now proving Proposition \ref{zero-condition}. , recalling there is only one nonzero entry in $[\bar A_\times, \bar b_\times]$.

{\bf Proof of Proposition \ref{zero-condition}:} { Sufficiency:} The sufficiency of condition a1) is obvious, as in this case for almost all $[A_*,b_*]\in {\bf S}_{[\bar A_*,\bar b_*]}$, ${\rm rank}([A_*,b_*][J_n,J_{n+1}\backslash \{j\}])=n$. Suppose condition a2) is fulfilled. We only need to consider the case where condition a1) is not fulfilled. In this case, as $(\bar A, \bar b)$ is structurally controllable, by Lemma \ref{structural-controllability-condition}, ${\rm grank}([\bar A, \bar b][J_n,J_{n+1}\backslash \{j\}])=n-1$ must hold. Then, for almost all $[A_*,b_*]\in {\bf S}_{[\bar A_*, \bar b_*]}$, ${\rm rank}([A_*,b_*][J_n,$ $J_{n+1}\backslash \{j\}])=n-1$. Next, consider two cases:
i) ${\rm grank}([\bar A, \bar b][J_n\backslash \{i\},$ $J_{n+1}\backslash \{j\}])=n-1$ and ii) ${\rm grank}([\bar A, \bar b][J_n\backslash \{i\},J_{n+1}\backslash \{j\}])<n-1$, i.e., ${\rm grank}([\bar A, \bar b][J_n\backslash \{i\},J_{n+1}\backslash \{j\}])=n-2$.  Let $q\in {\mathbb C}^{n}$ $(q\ne 0)$ be in the left null space of $[A_*,b_*][J_n,J_{n+1}\backslash \{j\}]$. Note $q$ is unique up to scaling. By Lemma \ref{null-space}, we generically have $q_k=0$ for each $k\in {\cal N}_*^j$; additionally, $q_i\ne 0$ in case i), and $q_i=0$ in case ii). Hence, in case i), for almost all $[A_*,b_*]\in {\bf S}_{[\bar A_*, \bar b_*]}$ and all $[A_\times, b_\times]\in {\bf S}_{[\bar A_\times, \bar b_\times]}$, it holds
$$ \begin{aligned}&q^\intercal[A_*+A_\times, b_*+b_\times][J_n,\{j\}]\\
&=\sum \nolimits_{k\in {\cal N}_*^j} q_k[A_*,b_*]_{kj}+ q_i[A_\times,b_\times]_{ij}
=q_i[A_\times,b_\times]_{ij}
\ne 0.\end{aligned}$$
In case ii), for almost all $[A_*,b_*]\in {\bf S}_{[\bar A_*, \bar b_*]}$ and {\emph{all}} $[A_\times, b_\times]\in {\bf S}_{[\bar A_\times, \bar b_\times]}$, we have
$$ \begin{aligned}&q^\intercal[A_*+A_\times, b_*+b_\times][J_n,\{j\}]
=\sum \nolimits_{k=1}^n q_k[A_*,b_*]_{kj}
=0,\end{aligned}$$
indicating $(A_*+A_\times, b_*+b_\times)$ is always uncontrollable, which is excluded by the structural controllability of $(\bar A, \bar b)$.

To show the sufficiency of condition a3), we only need to consider the case where condition a1) is not satisfied. In this case, for almost all $[A_*,b_*]\in {\bf S}_{[\bar A_*, \bar b_*]}$, following the similar arguments to those for condition a2), upon letting $q$ be in the left null space of $[A_*,b_*][J_n,J_{n+1}\backslash \{j\}]$, we have $q_i=0$, for almost all $[A_*,b_*]\in {\bf S}_{[\bar A_*, \bar b_*]}$. Then, for all $[A_\times, b_\times]\in {\bf S}_{[\bar A_\times, \bar b_\times]}$, it holds
$$ \begin{aligned}&q^\intercal[A_*+A_\times, b_*+b_\times][J_n,\{j\}]=\sum \nolimits_{k\in J_n,k\ne i} q_k[A_*,b_*]_{kj}
\myineqa 0,\end{aligned}$$
where (a) is due to the structural controllability of $(\bar A, \bar b)$.

{Necessity:} Suppose none of the three conditions is satisfied. Then, by the structural controllability of $(\bar A, \bar b)$ and Lemma \ref{structural-controllability-condition}, it must hold that ${\rm grank}([\bar A, \bar b][J_n,J_{n+1}\backslash \{j\}])={\rm grank}([\bar A, \bar b][J_n\backslash \{i\},$ $J_{n+1}\backslash \{j\}])=n-1$,  and there exists at least one $k\in J_n\backslash \{i\}$ so that $[\bar A,\bar b]_{kj}=*$ and ${\rm grank}([\bar A,\bar b][J_n\backslash \{k\},$ $J_{n+1}\backslash \{j\}])=n-1$. Under these conditions, for almost all $[A_*,b_*]\in {\bf S}_{[\bar A_*,\bar b_*]}$, we have ${\rm rank}([A_*,b_*][J_n,J_{n+1}\backslash \{j\}])=n-1$. Upon letting $q$ be a nonzero vector in the left null space of $[A_*,b_*][J_n,J_{n+1}\backslash \{j\}]$, Lemma \ref{null-space} yields that $q_i\ne 0$ and $q_k\ne 0$ generically hold. Assign
\begin{equation}\label{uncontrollable-value}[A_\times, b_\times]_{ij}=-1/{q_i}\sum \nolimits_{k\in J_n,k\ne i} q_k[A_*,b_*]_{kj}\ne 0,\end{equation}
where the inequality is due to the fact $[A_*,b_*]_{kj}$ is independent of $q$ ($q$ is uniquely determined by $[A_*,b_*][J_n,J_{n+1}\backslash \{j\}]$ up to scaling). We therefore have $$\begin{aligned}&q^\intercal[A_*+A_\times, b_*+b_\times][J_n,\{j\}]\\
&=\sum \nolimits_{k\in J_n,k\ne i} q_k[A_*,b_*]_{kj}+ q_i[A_\times,b_\times]_{ij}
=0.\end{aligned}$$ This proves the necessity.

{\bf Proof of Proposition \ref{graph-zero-mode}}: The equivalence is obvious since the generic rank of a structured matrix equals the size of a maximum matching of its associated bipartite graph (see Section \ref{sub-preli}).

{\bf Proof of Lemma \ref{row-rank-deficient}}: To prove Lemma \ref{row-rank-deficient}, we need the following auxiliary results.

\begin{lemma}[Lemma 9 of \cite{full-version-tac}]\label{root-independent}  Let $M$ be an $n\times n$ generic matrix over the variables $t_1,...,t_r$, and $E$ be an $n\times n$ constant matrix whose entries are either $0$ or $1$, where each row of $E$, as well as each column, has at most one $1$. Let ${\cal B}(P_\lambda)$ be the bipartite graph associated with the generic matrix pencil $P_\lambda\doteq M-\lambda E$ (defined in a way similar to ${\cal B}(H_\lambda)$). Let ${\cal T}_i$ be the set of variables of $t_1,...,t_r$ that appear in the $i$th column of $P_\lambda$. If ${\cal B}(P_\lambda)$ is DM-irreducible, then every nonzero root of $\det (P_\lambda)$ (if exists) cannot be independent of ${\cal T}_i$ (if nonempty), for each $i\in \{1,...,n\}$.
\end{lemma}

%The following lemma generalizes \citep[Lemma 10]{full-version-tac}. % The proof follows similar spirits to \citep[Lemma 10]{full-version-tac}.

{\bf Proof of Lemma \ref{row-rank-deficient} counting:} Remember for any $v_l^+\in V^+$, ${\cal B}(H_\lambda^{j_c})-\{v^+_l\}$ has a maximum matching with size $n-1$, as otherwise it cannot hold that ${\rm mt}({\cal B}(H_\lambda^{j_c}))=n$. We first prove property 1), and then property 2).   %Indeed, assuming $v_l^+\in V_{l'}^+$, $1\le l' \le d$, we have ${\rm mt}(G_k^{j_c})=|V_k^+|$ for $k\in \{1,...,d\}\backslash \{l\'\}$ and ${\rm mt}(G_{l'}^{j_c}-\{v_l^+\})=|V^+_{l'}|-1$, where the latter is due to the DM-irre. %In the following, by saying $M_{\lambda}^{j_c}$ (or its sub-matrices) satisfies certain properties, we mean these properties are satisfied for almost all values of the indeterminate parameters in $M_{\lambda}^{j_c}$ (or its sub-matrices).

{\bf Necessity in property 1):}  Suppose $\Omega_j^l\ne \emptyset$. Then, there is some $k'\in \Omega_j$, so that a matching $\cal M$ of ${\cal B}(H_\lambda^{j_c})-\{v_l^{+}\}$ with size $n-1$ exists that does not cover ${\cal G}^{j_c}_{k'}$. Suppose $v^-_{k''}\in V^-_{k'}$ is not matched by ${\cal M}\cap E_{k'}$.  As $k'\in \Omega_j$, Lemma \ref{nonzero-function} yields $\det M_{k'}^{j_c}(\lambda)$ generically has nonzero roots for $\lambda$, and let $z$ be one of such roots. Due to the block-triangular structure of $M_\lambda^{j_c}$, $z$ satisfies $\det M_z^{j_c}=0$. Note as ${\cal G}_{k'}^{j_c}$ is DM-irreducible, from Lemma \ref{root-independent}, $z$ cannot be independent of the indeterminate parameters in the column of $M^{j_c}_{k'}(\lambda)$ corresponding to $v_{k''}^-$  (note if $|V^+_{k'}|=1$ and $M^{j_c}_{k'}(\lambda)=\lambda$, then $k'\notin \Omega_j$). %; if $|V^+_{k'}|>1$ but the column of $M^{j_c}_{k'}(\lambda)$ corresponding to $v_{k''}^-$ contains only the variable $\lambda$, then ${\cal G}_{k'}^{j_c}$ is DM-reducible

On the other hand, $M_{t}^{j_c}(\lambda)$ for any $t< k'$ does not share a common factor with $M_{k'}^{j_c}(\lambda)$ except for the power of $\lambda$. By \citep[Lem 2]{Rational_function}, this implies $\det M_{t}^{j_c}(\lambda)$ for any $t< k'$ does not share a common nonzero root for $\lambda$ with $\det M_{k'}^{j_c}(\lambda)$, generically. Therefore, denoting $V^\star_{k'd}=V^\star_{k'}\cup V^\star_{k'+1}\cup \cdots V^\star_{d}$ where $\star=+,-$, if $M_z^{j_c}[V^+_{k'd}\backslash \{v_l^+\}, V^-_{k'd}]$ is of full row rank, then $M_z^{j_c}[V^+\backslash \{v_l^+\}, V^-]$ will be (due to its block-triangular structure). Noting \\ $\det (M_\lambda^{j_c}[V^+_{k'd}\backslash \{v^+_l\}, V^-_{k'd}\backslash \{v^-_{k''}\}])$ is not identically zero (because of the existence of $\cal M$; note $v_l^+\in V^+_{k'd}$, as otherwise every maximum matching of ${\cal B}(H_\lambda^{j_c})-\{v_l^+\}$ will cover ${\cal G}^{j_c}_{k'}$), any nonzero root of $\det (M_\lambda^{j_c}[V^+_{k'd}\backslash \{v^+_l\}, V^-_{k'd}\backslash \{v^-_{k''}\}])$ is independent of the parameters in the column of $M_{k'}^{j_c}(\lambda)$ corresponding to $v_{k''}^-$. This means,  $M_z^{j_c}[V^+_{k'd}\backslash \{v_l^+\}, V^-_{k'd}]$ is generically of full row rank, so is $M_z^{j_c}[V^+\backslash \{v^+_l\}, V^-]$.

{\bf Sufficiency in property 1) :} Suppose $\Omega_j^l=\emptyset$. Due to the block-triangular structure of $M_\lambda^{j_c}$ and from Lemma \ref{nonzero-function}, any nonzero root of $\det M_\lambda^{j_c}$, denoted by $z$, must be a nonzero root of $\det M_{k'}^{j_c}(\lambda)$ for some $k'\in \Omega_j$. Consider an arbitrary maximum matching $\cal M$ of ${\cal B}(H_\lambda^{j_c})-\{v_l^+\}$. Upon letting $V^-_{\cal M}$ be the set of vertices in $V^-$ that are matched by $\cal M$, we have $\bigcup \nolimits_{k\in \Omega_j} V^-_k\subseteq V^-_{\cal M}$. Let $R_1\doteq \bigcup\nolimits_{k\in \Omega_j}V_k^+$, $R_2\doteq V^+\backslash \{v_l^+\}$.   From \citep[Prop 2.1.3]{Murota_Book},
\begin{equation}\label{determinant-expression} {\footnotesize \begin{aligned}
&\det M_z^{j_c}[R_2, {V}^-_{\cal M}]=\\
&\sum \limits_{J\subseteq V^-_{\cal M},|J|=|R_1|} {\rm sgn}(R_1,J)\cdot {\det}M_z^{j_c}[R_1,J]\cdot \det M_z^{j_c}[R_2\backslash R_1,V^-_{\cal M}\backslash J],
\end{aligned}}
\end{equation}
where ${\rm sgn}(R_1,J)=\pm 1$ is the signature associated with $(R_1,J)$. Since each maximum matching of ${\cal B}(H_\lambda^{j_c})-\{v_l^+\}$ covers $\bigcup \nolimits_{k\in \Omega_j} {\cal G}_k^{j_c}$, ${\det}M_\lambda^{j_c}[R_1,J]\cdot \det M_\lambda^{j_c}[R_2\backslash R_1,V^-_{\cal M}\backslash J]$ $\equiv 0$ for any $J\subseteq V_{\cal M}^-$, $|J|=|R_1|$ but $J\ne \bigcup \nolimits_{k\in \Omega_j}V_k^-$, as otherwise $E(R_1,J)\cup E(R_2\backslash R_1,$ $V^-_{\cal M}\backslash J)$ will contain a maximum matching of ${\cal B}(H_\lambda^{j_c})-\{v_l^+\}$ that does not cover $\bigcup \nolimits_{k\in \Omega_j} {\cal G}_k^{j_c}$, meaning $\Omega_j^l\ne \emptyset$, where $E(V^+_s,V^-_s)$ denotes the set of edges between $V^+_s$ and $V^-_s$ in ${\cal B}(H_\lambda^{j_c})$, $V^+_s\subseteq V^+$, $V^-_s\subseteq V^-$. Note $\det M_z^{j_c}[R_1,\bigcup \nolimits_{k\in \Omega_j}V_k^-]=$\\ $\pm \prod \nolimits_{k\in \Omega_j}\det M_k^{j_c}(z)=0$ due to the block-triangular structure of $M_\lambda^{j_c}$ and $\det M_{k'}^{j_c}(z)=0$. Hence, $\det M_z^{j_c}[R_2, {V}^-_{\cal M}]=0$. Since this property holds for any maximum matching of ${\cal B}(H_\lambda^{j_c})-v_l^+$, $M_z^{j_c}[V^+\backslash \{v_l^+\}, V^-]$ is row rank deficient. %
% Observe that $M_\lambda^{j_c}[R_2, V^-({\cal M})]$ is also block-triangular with one of its diagonal blocks being $M_{k'}^{j_c}(\lambda)$. As $\det (M_{k'}^{j_c}(z))=0$, we have $\det(M_z^{j_c}[R_2, V^-({\cal M})])=0$.We are now proving property 2).

 {\bf Necessity in property 2):} Suppose there is a maximum matching of $B(H_\lambda^{j_c})-\{v_l^+\}$ that does not cover ${\cal G}_{k'}^{j_c}$ for some $k'\in \Omega_j^i$ (thus does not cover $\bigcup \nolimits_{k\in \Omega_j^i} {\cal G}_k^{j_c}$). By the definition of $\Omega_j^i$, there exists a maximum matching of $B(H_\lambda^{j_c})-\{v_i^+\}$ that does not cover ${\cal G}^{j_c}_{k'}$. In this case, following the similar argument to the proof of necessity in property 1), any nonzero root of $\det M_{k'}^{j_c}(\lambda)$, denoted by $z$, is also a root of $\det M_\lambda^{j_c}$, while making $M_{z}^{j_c}[V^+\backslash \{v^+_i\}, V^-]$ and $M_{z}^{j_c}[V^+\backslash \{v^+_l\}, V^-]$ of full row rank simultaneously.

{\bf Sufficiency in property 2):} From the proof of property 1), we know a nonzero $z$ makes $M_{z}^{j_c}[V^+\backslash \{v^+_i\}, V^-]$ of full row rank while $\det M_z^{j_c}=0$, if and only if $z$ is a nonzero root of $\det M^{j_c}_{k'}(\lambda)$ for some $k'\in \Omega_j$ satisfying the property that, ${\cal G}_{k'}^{j_c}$ is not covered by some maximum matching of $B(H_\lambda^{j_c})-\{v_i^+\}$ (referred to as condition f)). Indeed, following the necessity in property 1), we know a $z$ satisfying condition f) is sufficient; and following the similar arguments to the proof for sufficiency in property 1), we know condition f) for $z$ is necessary. By definition, the set of such a $k'$ satisfying condition f) is exactly $\Omega_j^i$. Again, following the similar manner to the proof for sufficiency in property 1) by replacing $M_z^{j_c}[R_1,J]$ with $M_z^{j_c}[\bigcup \nolimits_{k\in \Omega_j^i} V_k^+,J]$ in (\ref{determinant-expression}), it turns out $M_{z}^{j_c}[V^+\backslash \{v^+_{l}\}, V^-]$ is row rank deficient if every maximum matching of $B(H_\lambda^{j_c})-\{v_l^+\}$ covers $\bigcup \nolimits_{k\in \Omega_j^i} {\cal G}_k^{j_c}$.
%Property 2) follows immediately from the proof for the necessity of property 1).

{\bf Proof of Proposition \ref{nonzero-condition}:} In the following, we consider the generic realization $(\tilde A, \tilde b)$ of $(\bar A, \bar b)$. Recall $H_\lambda=[\tilde A-\lambda I, \tilde b]$. Let $\bar p_{ij} (\ne 0)$ be the parameter in the $(i,j)$th entry of $[\tilde A, \tilde b]$. For $v_l^+\in {V^+}$, suppose $v_l^+$ corresponds to the $\sigma(l)$th row of $M_\lambda^{j_c}$ after the row permutation $P$ on $H_\lambda^{j_c}$. %By saying $M_{\lambda}^{j_c}$ ($H_\lambda^{j_c}$ or its sub-matrices) satisfies certain properties, we mean these properties are satisfied for almost all values of the corresponding indeterminate parameters.

%by saying $M_{\lambda}^{j_c}$ (or its sub-matrices) satisfies certain properties, we mean these properties are satisfied for almost all values of the indeterminate parameters in $M_{\lambda}^{j_c}$ (or its sub-matrices) , with $P^{-1}$ and $Q^{-1}$ being permutation matrices

{\bf Sufficiency:} %Suppose for almost all $[A_*,b_*]\in {\bf S}_{[\bar A_*,\bar b_*]}$, such $[A_*,b_*]$, $z$ and $q$ that satisfy the requirements exist. Then, it follows $q^{\intercal}[A_*-zI,b_*][J_n,J_{n+1}\backslash \{j\}]=0$. Considering the generic realization $(\tilde A, \tilde b)$, this means $H^{j_c}_\lambda$ is row rank deficient at some nonzero $\lambda=z$.
Suppose condition c1) is fulfilled. If $\Omega_j=\emptyset$, by Lemma \ref{nonzero-function}, there is generically no $z\in {\mathbb C}\backslash \{0\}$ making $M_z^{j_c}$ row rank deficient, due to the block-triangular structure of $M_\lambda^{j_c}$. Noting $H_\lambda^{j_c}=P^{-1}M_{\lambda}^{j_c}Q^{-1}$ from (\ref{DM-decomp}), $H_z^{j_c}$ is generically of full rank for all nonzero $z$. This means $\Omega_j= \emptyset$ is sufficient. Now consider $\Omega_j\ne \emptyset$ but $\Omega^i_j=\emptyset$. In this case, consider an arbitrary $z \in {\mathbb C\backslash \{0\}}$ that makes $H^{j_c}_z$ row rank deficient. As $(\bar A, \bar b)$ is structurally controllable, $H_z^{j_c}$ generically has rank $n-1$ (otherwise $[H_z^{j_c},H_z^j]$ will be row rank deficient). Hence, upon letting $q$ be a nonzero vector in the left null space of $H_z^{j_c}$, $q$ is unique up to scaling. From property 1) of Lemma \ref{row-rank-deficient}, $H_z^{j_c}[V^+\backslash \{v_i^+\}, V^-]$ is row rank deficient (recalling $H_\lambda^{j_c}=P^{-1}M_{\lambda}^{j_c}Q^{-1}$). By Lemma \ref{null-space}, we have $q_i=0$. Hence, for any nonzero value of $\bar p_{ij}$,
$$q^\intercal H_z^j= \sum \nolimits_{k=1,k\ne i}^n q_k[H_z^j]_k\myineqa 0,$$
where $(a)$ is due to the controllability of $(\tilde A, \tilde b)$ and that $q^\intercal H_z^j$ is independent of $\bar p_{ij}$. Therefore, condition c1) is sufficient.

Suppose condition c2) is fulfilled. In this case, for those nonzero roots $z$ of $\det H_\lambda^{j_c}$ that make $H_z^{j_c}[V^+\backslash \{v_i^+\}, V^-]$ row rank deficient, following a similar argument to the above, we can obtain that there is no $\bar p_{ij}$ and $q$ ($\bar p_{ij}\ne 0$ and $q\in {\mathbb C}^n\backslash \{0\}$) making $q^\intercal H_z^j=0$. For
the nonzero root $z$ of $\det H_\lambda^{j_c}$ that makes $H_z^{j_c}[V^+\backslash \{v_i^+\}, V^-]$ of full row rank, let $q$ be a nonzero vector in the left null space of $H_z^{j_c}$ ($q$ is unique up to scaling). Then, according to property 2) of Lemma \ref{row-rank-deficient}, it turns out that
$q_i\ne 0$ and $q_k=0$ for all $k\in \{l: [H_z^j]_{l}\ne 0\}\backslash \{i\}$. We therefore have
$$q^\intercal H_z^j= \sum \nolimits_{k=1}^n q_k[H_z^j]_k\myeqa q_i\bar p_{ij}\ne 0,$$
where (a) is due to $[H_z^j]_i=\bar p_{ij}$ as $j\ne i$. Hence, condition c2) is also sufficient.

{\bf Necessity:} Suppose neither condition c1), nor condition c2) holds. Then, either i): $\Omega_j^i\ne \emptyset$, and the third item of condition c2) does not hold, or  ii): $\Omega_j^i\ne \emptyset$,  $i=j$, and the third item of condition 2) holds.  In case i), suppose the third item of condition 2) does not hold for a vertex $v_{l'}^+ \in {\cal N}({\cal B}(H_\lambda), v_j^-)\backslash \{v_i^+\}$. Considering the generic realization $(\tilde A, \tilde b)$, by property 2) of Lemma \ref{row-rank-deficient}, we known there is some nonzero $z$ making $M_z^{j_c}$ row rank deficient, while $M_{z}^{j_c}[V^+\backslash \{v^+_i\}, V^-]$ and $M_{z}^{j_c}[V^+\backslash \{v^+_{l'}\}, V^-]$ are of full row rank. Note it generically holds ${\rm rank}(M_z^{j_c})=n-1$, as otherwise ${\rm rank}([M_z^{j_c},M_z^j])<n$, contradicting the structural controllability of $(\bar A, \bar b)$. Let $\hat q$ be a nonzero vector spanning the left null space of $M_z^{j_c}$. Then, by Lemma \ref{null-space}, $\hat q_{\sigma(i)}\ne 0$ and $\hat q_{\sigma(l')}\ne 0$. Upon letting $[M_z^j]_{\sigma(i)}=-1/\hat q_{\sigma(i)}\sum \nolimits_{k=1,k\ne \sigma(i)}^n \hat q_k[M_z^j]_k$, we have
$$\hat qM_z^j=\hat q_{\sigma(i)}[M_z^j]_{\sigma(i)}+\sum \nolimits_{k=1,k\ne \sigma(i)}^n \hat q_{k}[M_z^j]_k =0,$$
which leads to $\hat qP[H_z^{j_c},H_z^j]=0$, by substituting $M_\lambda^{j_c}=PH_\lambda^{j_c}Q$, $M^j_\lambda=PH_\lambda^{j_c}$ and noting $Q$ is invertible. That is,
by assigning
\begin{equation}\label{uncontrollable-nz} \bar p_{ij}=
\begin{cases} z-1/\hat q_{\sigma(i)}\sum \nolimits_{k=1,k\ne \sigma(i)}^n \hat q_k[M_z^j]_k & \text{if}\ i=j \\
-1/\hat q_{\sigma(i)}\sum \nolimits_{k=1,k\ne \sigma(i)}^n \hat q_k[M_z^j]_k  &  \text{if}\ i\ne j,
\end{cases}\end{equation} we can make $\hat qP[\tilde A-zI,\tilde b]=0$. Note if $i=j$, $[M_z^j]_{\sigma(l')}$ is independent of $\hat q$ and $z$ (as $[M_z^j]_{\sigma(l')}$ contains a free parameter); if $i\ne j$, in case $\sum \nolimits_{k=1,k\ne \sigma(i)}^n \hat q_k[M_z^j]_k$ contains more than one nonzero items, at least one $[M_z^j]_k$ is independent of $\hat q$ and $z$. Hence, in all these circumstances, it is assured $\bar p_{ij}\ne 0$.

As for case ii), following similar arguments to case i), upon letting $\hat q\ne 0$ be in the left null space of $M_z^{j_c}$ for some nonzero $z$ making $\det  M_z^{j_c}=0$ while $M_z^{j_c}[V^+\backslash \{v_i^+\},V^-]$ of full row rank, we have $\hat q_{\sigma(i)}\ne 0$ and $\hat q_{\sigma(k)}=0$ for all $k$ satisfying $v_k^+\in {\cal N}({\cal B}(H_\lambda), v_j^-)\backslash \{v_i^+\}$. Hence, if $p_{ii}=z\ne 0$, then $\hat q M_z^j= \hat q_{\sigma(i)}(p_{ii}-z)=0$, leading to $\hat q P[H_z^{j_c},H_z^j]\!=\!0$.
Since $(\tilde A, \tilde b)$ is a generic realization of $(\bar A, \bar b)$, in both cases, we know for almost all $[A_*, b_*]\in {\bf S}_{[\bar A_*, \bar b_*]}$, there exist $[A_\times, b_\times]\in {\bf S}_{[\bar A_\times, \bar b_\times]}$, nonzero $z$ and $q\in {\mathbb C}^n$ satisfying $q^\intercal [A_*+A_\times-z I, b_*+b_\times]=0$.

{\bf Proof of Theorem \ref{main-result}}: The result comes immediately from the PBH test and Propositions \ref{one-edge-principle}, \ref{graph-zero-mode} and \ref{nonzero-condition}.

{\bf{Proof of Proposition \ref{pro-generic-multi}:}} The case where $(\bar A, \bar B)$ is not structurally controllable is trivial. Now assume structural controllability of $(\bar A, \bar B)$, and consider its generic realization $(\tilde A, \tilde B)$. Let ${\bar p}$ be the parameter for the unique $\times$ entry in $[\tilde A, \tilde B]$, and $p$ the collection of parameters for the remaining indeterminate entries. Let $\Gamma(p,{\bar p})$ be the greatest common divisor among all determinants of the $n\times n$ submatrices of ${\cal C}(\tilde A, \tilde B)$. According to \citep[Lem 2]{Rational_function}, for almost all $p\in {\mathbb C}^{n_*}$ (recalling $n^*$ is the number of $*$ entries), the determinants of all the $n\times n$ submatrices of ${\cal C}(\tilde A, \tilde B)$ share a common zero for ${\bar p}$, if and only if the leading degree for ${\bar p}$ in $\Gamma(p,{\bar p})$ is no less than one. Therefore, if $\Gamma(p,{\bar p})=P(p){\bar p}^r$, where $r\ge 0$ and $P(p)$ is a nonzero polynomial of $p$, then for almost all $p\in {\mathbb C}^n$ satisfying $P(p)\ne 0$ and all ${\bar p}\ne 0$,
${\cal C}(\tilde A, \tilde B)$ is of full row rank, indicating that the corresponding realization is controllable. Otherwise, if $\Gamma(p,{\bar p})$ contains two monomials with different degrees for ${\bar p}$, then for almost all $p\in {\mathbb C}^{n_*}$, there exists a nonzero solution ${\bar p}$ satisfying $\Gamma(p,{\bar p})=0$ (see the proof of Proposition \ref{algebraic-condition}), making the corresponding realization uncontrollable.

{\bf Proof of Proposition \ref{zero-mode-multi}}: Note Lemma \ref{null-space} is applicable to rectangular matrices. This means the proof can be completed in the same manner as in Propositions \ref{zero-condition} and \ref{graph-zero-mode}, which thus is omitted.

%{\bf Proof of Proposition \ref{nonzero-mode-multi}:}  To prove this proposition, we need the following auxiliary result.
%\begin{lemma} \label{horizontal-no-root} If $m>1$, there is generically no nonzero $\lambda$ that can make $H^{j_c}_\lambda[V^+_0,V^-_0]$ row rank deficient.
%\end{lemma} {\bf Proof of Proposition \ref{nonzero-mode-multi}:}  To prove this proposition, we need the following auxiliary result. ({\bf Proof continuing})

{\bf Proof of Lemma \ref{horizontal-no-root}}: Since ${\cal G}_0^{j_c}$ is the horizontal tail, from \cite[Corollary 2.2.23]{Murota_Book}, for each $v_l^-\in V_0^-$, ${\rm mt}({\cal G}_0^{j_c}-\{v_l^-\})=|V_0^+|$. From \citep[Lem 9]{full-version-tac}, for all $v_l^-\in V_0^-$, every maximum matching of ${\cal G}_0^{j_c}$ corresponds to a nonzero term in the determinant of a $|V_0^+|\times |V_0^+|$ submatrix of $H^{j_c}_\lambda[V^+_0,V^-_0]$ that cannot be cancelled out by other terms. Therefore, supposing there is a nonzero value, denoted by $z$, such that $H^{j_c}_z[V^+_0,V^-_0]$ is of row rank deficient, $z$ should depend only on the free parameters in $H^{j_c}_{\lambda}[V^+_0,V^-_0\backslash \{v_l^-\}]$, $\forall v_l^-\in V_0^-$. Applying this across $v_l^-\in V_0^-$, it turns out that $z$ is independent of the free parameters in each column of $H_\lambda^{j_c}$, causing a contradiction.
%\end{proof}

{\bf Proof of Proposition \ref{nonzero-mode-multi}:} From Lemmas \ref{nonzero-function} and \ref{horizontal-no-root}, if $z$ is a nonzero root of $\det H_\lambda^{j_c}[V_k^+,V_k^-]$ for some $k\in \Omega_j$, then $z$ also makes $H_\lambda^{j_c}$ row rank deficient owing to the block-triangular structure of its DM-decomposition; and vice versa. With these results,
it can be proved easily that, a nonzero $z$ makes \\$H_\lambda^{j_c}[V^+\backslash \{v_l^+\}, V^-]$ row rank deficient for a given $v_l^+\in V^+$, if and only if $z$ makes $H_\lambda^{j_c}[V^+\backslash (V^+_0\cup \{v_l^+\}), V^-\backslash V^-_0]$ row rank deficient. Having observed this, it can be found Lemma \ref{row-rank-deficient} still holds for the rectangular matrix $H_\lambda^{j_c}$ (by changing $\det M_\lambda^{j_c}=0$ to the row-rank deficient of $H_\lambda^{j_c}$). Hence, the proof can be completed in the similar manner to that for the single-input case, i.e., the proof for Proposition \ref{nonzero-condition}. The details are omitted due to their similarities.

{\bf Proof of Corollary \ref{SSC-corollary}:} Sufficiency: From Theorems \ref{algebraic-condition} and \ref{main-result}, we know $\det {\cal C}(\tilde A, \tilde b)$ has the form $\det {\cal C}(\tilde A, \tilde b)= \prod \nolimits_{i=1}^{n_\times} \bar p_i^{r_i}$, $r_i\ge 0$ under the proposed conditions,  where $[\tilde A, \tilde b]$ is a generic realization of $[\bar A, \bar b]$ with the indeterminate parameters being $\bar p_1,...,\bar p_{n_\times}$. It is then obvious that for all $(\bar p_1,...,\bar p_{n_\times})\in \bar {\mathbb R}^{n_\times}$, the corresponding system realization is controllable.

Necessity: We prove the necessity by contradiction. Let $\bar p_\pi$ be the parameter for the $\pi$th $\times$ entry of $[\bar A, \bar b]$ for a $\pi\in {\cal N}_{\times}$, and $\bar p_{\pi^{c}}$ the vector consisting of the parameters for the remaining $\times$ entries (except $\bar p_{\pi}$). The necessity of structural controllability of $(\bar A, \bar b)$ is obvious. Now assume $(\bar A, \bar b)$ is structurally controllable.   Suppose i) is not fulfilled for a $\pi=(i,j)\in {\cal N}_{\times}$.  From the proof for necessity of Proposition \ref{zero-condition}, for almost all $\bar p_{\pi^{c}}\in \bar {\mathbb R}^{n_\times-1}$, there exists nonzero real $\bar p_\pi$ (expressed in (\ref{uncontrollable-value}), where $q$ is chosen to be real), such that the corresponding real system realization is uncontrollable. This means $(\bar A, \bar b)$ is not SSC in the real field.

% $[\bar A^\pi_1, \bar b^\pi_1]$ be obtained from $[\bar A, \bar b]$ by replacing its $\pi$th $\times$ entry with fixed zero, and $[\bar A^\pi_2,\bar b^\pi_2]$ be obtained from $[\bar A, \bar b]$ by preserving only its $\pi$th $\times$ entry and replacing the rest with fixed zero.

Furthermore, suppose ii) is not fulfilled for a $\pi=(i,j)\in {\cal N}_{\times}$. We first consider the case $\Omega_j^i\ne \emptyset$ and the third item of condition c2) does not hold for some $v_{l'}^+ \in N({\cal B}(H_\lambda), v_j^-)\backslash \{v_i^+\}$. Following the proof for necessity of Proposition \ref{nonzero-condition}, in this case, to construct a real uncontrollable realization of $(\bar A, \bar b)$, we only need to demonstrate $z$ in (\ref{uncontrollable-nz}) can be real (thus the vector $\hat q$ therein can be real too). To this end, recall in (\ref{uncontrollable-nz}), $z$ needs to satisfy: requirement 1) $z$ makes $M_z^{j_c}$ row rank deficient, and requirement 2) $z$ makes $M_z^{j_c}[V^+\backslash \{v_i^+\}, V^-]$ and $M_z^{j_c}[V^+\backslash \{v_{l'}^+\},V^-]$ of full row rank. Notice from the proof for necessity of Lemma \ref{row-rank-deficient}, if $z$ is a real nonzero root of $\det M_{k'}^{j_c}(\lambda)$ for some $k'$, where $k'\in \Omega_j^i$ is such that there is a maximum matching ${\cal M}_2$ of ${\cal B}(H_\lambda^{j_c})-\{v_{l'}^+\}$ not covering ${\cal G}^{j_c}_{k'}$, then $z$ generically satisfies the previous two requirements. Let us assume that vertex $v^-_{j_2}$ of ${\cal G}^{j_c}_{k'}$ is not matched by ${\cal M}_2\cap E_{k'}$, and $\pi_2=(i_2,j_2)$ corresponds to a nonzero entry of $M_{k'}^{j_c}(\lambda)$.  As $k'\in \Omega_j^i$, there is a maximum matching ${\cal M}_1$ of ${\cal B}(H_\lambda^{j_c})-\{v_i^+\}$ not covering ${\cal G}^{j_c}_{k'}$. Similarly, assume that vertex $v^-_{j_1}$ of ${\cal G}^{j_c}_{k'}$ is not matched by ${\cal M}_1\cap E_{k'}$, and $\pi_1=(i_1,j_1)$ corresponds to a nonzero entry of $M_{k'}^{j_c}(\lambda)$. Let $R({\cal M})$ and $C({\cal M})$ be respectively the set of vertices in $V^+$ and in $V^-$ that are matched by a matching ${\cal M}$. Denote $V^\star_{1(k'-1)}=V^\star_1\cup\cdots V^\star_{k'-1}$ and $V^\star_{k'd}=V^\star_{k'}\cup \cdots V^{\star}_{d}$, where $\star=+$ or $-$.

%A multi-affine polynomial over a field is either constant or takes all values in the field \citep[Lemma 2.7]{buss1999computational}

Since ${\cal G}^{j_c}_{k'}$ is DM-irreducible, by Lemma \ref{root-independent}, $\det M_{k'}^{j_c}(\lambda)$ does not have a fixed nonzero root for $\lambda$ as $\bar p_{\pi^{c}}$ varies. Note also for any {\emph{fixed}} $z\in \bar {\mathbb R}$, $\det M_{k'}^{j_c}(z)$ is both irreducible (i.e., cannot be factored as two polynomials with smaller degrees) and multi-affine in $\bar p_{\pi^{c}}$ (i.e., every variable in $\bar p_{\pi^{c}}$  occurs with degree $0$ or $1$ in every term). Hence, $\det M_{k'}^{j_c}(\lambda)$ can be written as $\bar p_{\pi_1}f_1(\lambda,\bar p_{\pi^{c'}_1})+ f_0(\lambda,\bar p_{\pi^{c'}_1})$, where $f_1(\lambda,\bar p_{\pi^{c'}_1})$ and $f_0(\lambda,\bar p_{\pi^{c'}_1})$ are polynomials of $\lambda$ and $\bar p_{\pi^{c'}_1}$, with $\bar p_{\pi^{c'}_1}$ being the collection of indeterminate parameters in $M_{k'}^{j_c}(\lambda)$ except $\bar p_{\pi_1}$. Based on the above arguments, the proof for necessity of Lemma \ref{row-rank-deficient} indicates that the following constraints suffice to satisfy requirements 1) and 2):
%\begin{subequations}
\begin{align}
   \bar p_{\pi_1}f_1(z,\bar p_{\pi^{c'}_1})+ f_0(z,\bar p_{\pi^{c'}_1})&=0 \label{subeq1}\\
    f_1(z,\bar p_{\pi^{c'}_1})&\ne 0 \label{subeq2} \\
    f_0(z,\bar p_{\pi^{c'}_1}) &\ne 0 \label{subeq3} \\
    \det M_z^{j_c}[R({\cal M}_1)\cap V^+_{k'd}, C({\cal M}_1)\cap V^-_{k'd}]&\ne 0 \label{subeq4} \\
    \det M_z^{j_c}[R({\cal M}_2)\cap V^+_{k'd}, C({\cal M}_2)\cap V^-_{k'd}]&\ne 0 \label{subeq5} \\
    \det M_z^{j_c}[V^+_{1(k'-1)},V^-_{l(k'-1)}]&\ne 0 \label{subeq6}.
\end{align}
Indeed, (\ref{subeq1}) ensures $\det M_{k'}^{j_c}(z)=0$ leading to $\det M_z^{j_c}=0$, while (\ref{subeq4}), (\ref{subeq5}) and (\ref{subeq6}) ensure that requirement 2) is fulfilled (due to the block-diagonal structure of $M_z^{j_c}$). Now consider $z$ and $\bar p_{\pi^{c}}\backslash \{\bar p_{\pi_1}\}$ as nonzero real numbers.  If $j_2=j_1$, the constraints (\ref{subeq2})-(\ref{subeq6}) do not not involve $\bar p_{\pi_1}$. Then, for almost all nonzero real numbers of $z$ and $\bar p_{\pi^c}\backslash \{\bar p_{\pi_1}\}$, there is a $\bar p_{\pi_{1}}\in {\bar {\mathbb R}}$ satisfying (\ref{subeq1})-(\ref{subeq6}). If $j_2\ne j_1$, then $\bar p_{\pi_1}$ satisfying (\ref{subeq1}) depends generically on $z$ and $\bar p_{\pi^{c'}_1}$ (by Lemma \ref{root-independent}), thus not independent of ${\bar p}_{\pi_2}$, while $\bar p_{\pi_1}$ satisfying (\ref{subeq5}) is independent of $\bar p_{\pi_2}$. This means, again, for almost all nonzero real numbers of $z$ and $\bar p_{\pi^c}\backslash \{\bar p_{\pi_1}\}$, there is $\bar p_{\pi_1}\in {\bar {\mathbb R}}$ satisfying (\ref{subeq1})-(\ref{subeq6}). Therefore, for almost all $z\in \bar {\mathbb R}$, we can always find $\bar p_{\pi^{c}}\in \bar {\mathbb R}^{n_\times-1}$ such that requirements 1) and 2) are satisfied. Then, after determining $\bar p_{\pi}$ according to (\ref{uncontrollable-nz}), where $\hat q$ and $z$ are both real, the corresponding real system realization will be uncontrollable, meaning that $(\bar A, \bar b)$ is not SSC in the real field. For the case where $\Omega_j^i\ne \emptyset$,  $i=j$, and the third item of condition 2) holds, we can adopt a similar argument to construct a real uncontrollable  realization of $(\bar A, \bar b)$. This proves the necessity.

}

\section*{\refname}
\bibliographystyle{elsarticle-num}
{\footnotesize
\bibliography{yuanz3}
}
%\end{reference}

\end{document}